\documentclass[11pt]{article}
\usepackage[left=2cm,top=2cm,right=2cm,bottom=2.5cm,foot=2cm, nohead]{geometry}
\usepackage{amsmath,amsfonts,amssymb,amsthm,mathrsfs}
\usepackage{color, graphicx}
\usepackage{epsfig}
\usepackage{natbib}
\bibpunct{[}{]}{,}{a}{}{;}
\usepackage{hyperref}
\usepackage{pdflscape}
\usepackage{lineno}
\usepackage{soul}
\usepackage[normalem]{ulem}
\usepackage{authblk}

\definecolor{OliveGreen}{rgb}{0,0.6,0}
\hypersetup{colorlinks=true, citecolor=OliveGreen}
\usepackage{multirow}
\usepackage{color}

\usepackage{algorithm}
\usepackage{algpseudocode}

\usepackage{arydshln}

\usepackage{setspace,enumitem}

\makeatletter
\def\BState{\State\hskip-\ALG@thistlm}
\makeatother

\newcommand{\argmin}{\operatornamewithlimits{argmin}}

\newcommand{\BEA}{\begin{eqnarray}}
\newcommand{\EEA}{\end{eqnarray}}
\newcommand{\BEQ}{\begin{equation}}
\newcommand{\EEQ}{\end{equation}}
\newcommand{\BIT}{\begin{itemize}}
\newcommand{\EIT}{\end{itemize}}
\newcommand{\BNUM}{\begin{enumerate}}
\newcommand{\ENUM}{\end{enumerate}}

\newcommand{\bs}{\boldsymbol}
\newcommand{\MC}{\mathcal}

\newcommand{\MBE}{\mathbb{E}}
\newcommand{\MBP}{\mathbb{P}}
\newcommand{\MBR}{\mathbb{R}}

\newcommand{\NN}{\nonumber}
\newcommand{\TD}{\tilde}
\newcommand{\WH}{\widehat}

\newtheorem{proposition}{Proposition}

\newtheorem{lemma}{Lemma}

\begin{document}

\title{Simultaneous disease mapping and hot spot detection  with  application to childhood obesity surveillance from electronic health records\thanks{This work was done while Choi was a postdoctoral fellow at Public Health Sciences Division, Fred Hutchinson Cancer Research Center.}}

%\title{Penalty-based spatial smoothing and outlier detection for childhood obesity surveillance from electronic health records}
%\title{Granular-level childhood obesity surveillance from electronic health records: simultaneous outlier detection and trend filtering in generalized linear model}
\date{\today}

\author[1]{Young-Geun Choi}
\author[2]{Lawrence P. Hanrahan}
\author[3]{Derek Norton}
\author[4]{Ying-Qi Zhao\thanks{Corresponding author. Email: \url{yqzhao@fredhutch.org}}}
\affil[1]{Data Labs, SK Telecom}
\affil[2]{Department of Family Medicine and Community Health, University of Wisconsin-Madison}
\affil[3]{Department of Biostatistics and Medical Informatics, University of Wisconsin-Madison}
\affil[4]{Public Health Sciences Division, Fred Hutchinson Cancer Research Center}
\maketitle

\begin{abstract}

Electronic health records (EHRs) have become a platform for data-driven surveillance on a granular level in recent years. In this paper, we make use of EHRs for early prevention of childhood obesity. 
The proposed method simultaneously provides smooth disease mapping and outlier information for  obesity prevalence, which are useful for raising public awareness and facilitating targeted intervention. 
More precisely, we consider a penalized multilevel generalized linear model. We decompose  regional contribution  into smooth and sparse signals, which are automatically identified by a combination of fusion and sparse penalties imposed on the likelihood function. In addition,  we weigh the proposed likelihood to account for the missingness and potential non-representativeness arising from the EHR data. We develop a novel alternating minimization algorithm, which is computationally efficient, easy to implement, and guarantees convergence. Simulation studies demonstrate superior performance of the proposed method. Finally, we apply our method to the University of Wisconsin Population Health Information Exchange database.
\newline

\end{abstract}

\baselineskip 16pt

\section{Introduction}

Childhood obesity prevention has become increasingly important to control the global obesity pandemic. Granular-level surveillance of childhood obesity that identifies and tracks obesity trends  is needed to  
 help  design interventions and guide policy solutions when  monetary resources are limited. \citep{Longjohn2010}.
%
%Most of the national datasets, such as the Nutrition Examination Survey, have limited utility in evaluating obesity at local levels. 
Several efforts have been made to construct local surveillance systems \citep{Hoelscher2017}, which are primarily school-based \citep{Blondin2016}. 
%For example, \cite{Blondin2016} reported that 14 states collect child BMI data  for public health surveillance, 13 of which are school-based. 
Although  school-based surveillance can be effective for data collection and implementation, there are concerns about privacy, stigmatization, and dysfunctional behavioral responses \citep{Mass2014}. Alternatively, routinely collected massive health databases such as Electronic Health Records (EHRs)  are gaining  attention as a platform for assessing trends and local childhood obesity risk \citep{Friedman2013}. %Lager2009, Tatem2017

Statistical methods for geospatial surveillance may include two aspects: i) monitoring regional trends in prevalence (also known as ``disease mapping'' or ``risk mapping'') and ii) identifying unexpected variation in the prevalence of different locations (also known as ``hot spot detection'').
Traditionally, these two tasks have been accomplished separately. 

For task i), 
obesity literature mainly utilized the standard generalized linear mixed effect model (GLMM) to account for individual factors and community environments \citep{Zhang2011, Davila-Payan2015}.  %Drewnowski2014,Lin2017
Those approaches assumed the regional random effects to be independent, although a spatial dependency exists even after adjusting for covariates \citep{Panczak2016}.  %Huang2015a,
%
%  
% continuous smoothing
To account for the spatial dependence, methods for smooth disease mapping  have been proposed from both frequentist and Bayesian perspectives.
Under Poisson log-linear models or multilevel logistic models, the region-specific effects were smoothed by kernels \citep{Ghosh1999} or splines \citep{Ugarte2010,Maiti2016}, or were modeled as a dependent random vector by conditional autoregressive (CAR) priors  \citep{Waller1997,Pascutto2000,Lee2013a,Mercer2015}.  
These strategies resulted in ``clustered'' risk maps, which enhanced interpretability, but did not explore identification of aberrant {regions}. 
For task ii), 
%
% ii) scan statistic, residual
%% scan statistic
the most popular approach for detecting locations with outbreaking incidence is the spatial scan statistics approach \citep{Kulldorff1995, Kulldorff1997, Jung2009}. %which was further extended to general outcomes \citep{Jung2007,Huang2007,Kulldorff2009} and  covariates adjustment \citep{Jung2009}. 
The scan statistic methods search over a pre-specified set of  geographical districts  and conduct  a generalized likelihood ratio test for testing whether the proportions of events are homogeneous across, inside, and outside the district.  %The method yields clusters of locations that reject the null, i.e., locations with the lowest or highest event rates.  %T refer to the books
%Although the scan statistics were successful in identifying multiple clusters with fixed sizes, 
 However, it may not be suitable for identifying multiple locations with heterogeneous sizes. Residuals generated from regression approaches can also be used to detect regional outbreaks \citep{Kafadar1992, Farrington1996, Zhao2011}.   
%The estimation of the residual variance in Poisson log-linear models was discussed in. 
However, residual-based outlier detection is  known to fail when an outlier is a leverage point or there are multiple outliers \citep{She2011}.

Use of the fusion penalty for smoothing was first proposed in a least squares setup \citep{Tibshirani2005}.  The resulting fit from the fusion penalty appears to be piecewise constant, yielding  a natural clustering of fitted values.  
Smoothing by the fusion penalty enables an additional regularization using a different penalty, such as a sparse penalty, which may not be straightforward in other smooth disease mapping methods.  
Sparse penalty for outlier detection was used  with  the squared error loss \citep{Kim2009, Tibshirani2011a, She2011}.  \cite{Kim2009} and \cite{Tibshirani2011a} considered the $\ell_1$ penalty, and \cite{She2011} reported that nonconvex penalties outperformed both $\ell_1$ penalty  and residual-based approaches  for  detection in  standard multiple linear regression.  % with varying magnitudes and numbers of outliers.

In this paper, we develop a new method  that simultaneously produces an interpretable disease map and {detects outlier regions}.
We formulate a multilevel logistic model to naturally incorporate risk factors. 
 A novel hybrid regularization is introduced, where the region-specific effect is represented by the summation of a smooth signal and a sparse signal.
The smooth signal is regularized by a fusion penalty so that adjacent locations  tend to have similar fitted baseline obesity rates.
A nonconvex sparse penalty is enforced for the sparse signals  so that nonzero fitted coefficients signify potential outliers. 
It is worthwhile mentioning that 
estimating population health metrics from EHRs can be particularly challenging. First, subjects  in EHRs systems are not randomly sampled. EHRs  only capture people seeking healthcare. % for various reasons, including visits for preventive services, disease management,  acute care etc.  
Second, recorded data often suffers from missingness, 
% \cite{Klompas2017}
 because for each patient, EHRs only collect data on  the tests and conditions that clinicians order  and diagnose. 
% what kind of approach?
% there are some success stories
Thus, statistical models utilizing EHRs should deal with  missingness and non-representativeness. 
%
%Some authors have demonstrated that EHRs incorporated with sample reweighting procedures can produce prevalence estimates  comparable to those from national survey data \citep{Flood2015, Klompas2017}. %McVeigh2016, 
Following \cite{Flood2015}, we adopt a two-step weighting procedure to account for missing data  and to adjust the covariate distribution for a nationally representative sample.
% add  grouping of geo- graphical units with similar risk
We develop an alternating minimization  algorithm  to optimize a nonconvex objective function, which is computationally efficient and can leverage off-the-shelf software packages.

% use of penalty can detect multiple outliers better
% 
Our original contributions are twofold. First, while the hybrid regularization of the fusion and $\ell_1$ penalties has been considered in  linear models \citep{Kim2009,Tibshirani2011a},  to the best of our knowledge, we are the first  to incorporate a fusion penalty and a nonconvex penalty to identify outliers.
%\sout{Second,  we implement  fusion penalty-based regression in a generalized linear model, which was primarily employed for  linear models.} \red{What do you mean by this? is fusion penalty-based regression usually employed in linear models, but not generalized linear models? (YG: my original point was that we provided algorithm with convergence guarantee, beyond linear models. I tried to express it as below.)} 
%
{Second,  we provide an efficient optimization algorithm that guarantees convergence for the hybrid regularization model.}
%\sout{We provide an efficient optimization algorithm, which can be easily extended to handle other convex loss functions.}  %\red{to deal with areal data with countable or continuous responses}.
{Although our algorithm is described in a Bernoulli likelihood, it can be easily extended to handle other convex loss functions.}

%The rest of the paper is organized as follows. 
In Section \ref{sec:data}, we introduce the University of Wisconsin Electronic Health Record Public Health Information Exchange (PHINEX) database  that motivated our study. 
We introduce our model and formalize the objective function in Section \ref{sec:method}.  We also develop a computational algorithm and discuss tuning parameter selection in this section.   
Simulation studies are presented in Section \ref{sec:simul}, which demonstrate the superior performance of our proposed method. 
We apply our method to  PHINEX on childhood obesity surveillance in Section \ref{sec:application}. We provide concluding remarks in Section \ref{sec:conclusion}.

\section{Data}\label{sec:data}

The University of Wisconsin Electronic Health Record Public Health Information Exchange (UW eHealth PHINEX) database contains EHR data from a south-central Wisconsin academic healthcare system. It consists of patient records with  documented primary care encounters at family medicine, pediatric, and internal medicine clinics occurring from 2007 to 2012. All PHINEX data were derived from the Epic EHR Clarity Database (EpicCare Electronic Medical Record, Epic Systems Corp., Verona WI). Furthermore, the program geocodes to the census blockgroup and links EHRs with community-level  social determinants of health. It was created to improve clinical practice and population health by understanding local variations in disease risk, patients, and communities \citep{guilbert2012}. %This paper focuses on developing novel methods for local community level childhood obesity surveillance using  EHR data. 

In this paper, we focused on 93,130 patients aged 2--19 years during 2011--2012. Body mass index (BMI) values (in $kg/m^2$) were calculated from a subject's height and weight, measured at the same visit.
%These values, along with the age / sex growth charts from the Centers for Disease Control and Prevention (CDC, year 2000), were used to assess the BMI percentile of the subject.
Any subject with a BMI at or above the 95th percentile was categorized as obese.  % \citep{Spear2007}.
  Among all the patients, 34,852 (37.4\%) were missing a valid BMI.   
Subject covariates included sex, age, race/ethnicity, health service payer (i.e., insurance), and the 2010 census blockgroup information on subject residence.  Economic hardship index (EHI) \citep{Nathan1989} was used as a measure of  blockgroup socioeconomic status, and was calculated from a blockgroup's: \% of housing units with more than one person per room; \% of households below the federal poverty level; \% of people \textgreater 16 years of age who are unemployed; \% of people \textgreater 25 years of age without a high school education; \% of people \textless 18 or \textgreater 64 years of age; and per capita income. EHI was normalized for all Wisconsin census blockgroups.
The values were continuous, ranging from 0 to 100, with larger values indicating greater hardship.
Urbanicity of a census blockgroup was based on  its  11 Urbanization Summary Groups, according to \cite{ESRI2012}. 
These groups were derived from data on census blockgroup population density, city size, proximity to metropolitan areas, and economic/social centrality.
Urbanicity integer values ranged from 1 (the most urban)  to 11 (the most rural).

\section{Method}\label{sec:method}

% Objective function
\subsection{Model setup}\label{subsec:model}

We use a double subscript, $ij$ ($j=1,\ldots,n_i$, $i=1,\ldots,K$) to indicate the $j$-th subject in the $i$-th region.  Let $\bs{S}_i$ be the location of the $i$-th region.  Let $\bs{X}_i$ denote the region-level covariates such as urbanicity and EHI. 
 Let $Y_{ij}$ be the obese indicator of the $(ij)$-th subject, with $Y_{ij} = 1$ indicating obese. Lastly, let $\bs{Z}_{ij}$ be a vector of the covariates of the $(ij)$-th subject such as gender, age, race/ethnicity and insurance payor.

Let $p_{ij} = \MBP(Y_{ij} = 1| \bs{Z}_{ij}, \bs{X}_i)$.  We formalize our model for the $p_{ij}$ as
\begin{gather}
{\rm logit}(p_{ij}) = 
	\bs{Z}_{ij}^T \bs{\alpha}_1 + \bs{X}_{i}^T \bs{\alpha}_2 + \beta_i + \gamma_i, 	
	\label{eqn:model} \\
% \sum_{(i_1, i_2): \|\bs{S}_{i_1} - \bs{S}_{i_2}\| \leq h}
\mbox{subject~to} \qquad	 \sum_{i_1 < i_2} \rho_{i_1,i_2} |\beta_{i_1} - \beta_{i_2}| \leq c_1 ; 
					\label{eqn:const1} \\
\qquad\qquad \sum_{i=1}^K I(\gamma_i \neq 0 ) \leq c_2,
	\label{eqn:const2}
\end{gather}
   where $c_1, c_2 \geq 0$, ${\rm logit}(t) = \log \{t / (1-t)\}$,  and $I(\cdot)$ is the indicator function. Here, $\beta_i$ represents a regional-specific effect for the $i$-th region  that is not explained by $\bs{X}_i$. Since the probability of a child being obese might be affected by the community environment  he or she resides in, we expect the regional contribution on the obesity prevalence to be similar for individuals in closer locations \citep{Panczak2016}, and thus a smoothness constraint $\eqref{eqn:const1}$ is imposed on $\beta_i$. %\red{Such spatial dependency has been discussed in the obesity literature \citep{Huang2015a,Panczak2016}.}
The fusion weight $\rho_{i_1,i_2}$ ($\rho_{i_1,i_2} \geq 0$) represents the strength of the ``fusion'' for each pair of $i_1$ and $i_2$. A higher value of $\rho_{i_1,i_2}$ will lead to a more similar pair of the fitted  $\beta_{i_1}$ and $\beta_{i_2}$. %  Note that although we use geodistance to define the similarity between different locations to define $\rho_{i_1,i_2}$, our model allows other measures of similarity depending on the context.  
With an appropriate choice of tuning parameter, the values that $\beta_i$ could take are limited, where similar locations are grouped together.  
We may interpret the distinct levels of $\beta_i$ as segmentation or clustering of the regions. We also note that \eqref{eqn:const1}  can be seen as a two-dimensional generalization of the total variation constraint used in the fused lasso  \citep{Tibshirani2005}. 
$\gamma_i$ is introduced to capture potential aberrant regions, where the $i$-th region  is an outlier with unusual obesity  prevalence if $\gamma_i \neq 0$. Given the sparsity constraint \eqref{eqn:const2},  we expect $\gamma_i$ will be zero (non-outlier) for most regions, but a few  might be nonzero (outliers).  

\subsection{Estimation with complete data}

Denote by $N = \sum_{i=1}^K n_i$ and define $\bs{\alpha} = (\bs{\alpha}_1^T, \bs{\alpha}_2^T)^T$, $\bs{\beta} = (\beta_1, \ldots, \beta_K)^T$ and $\bs{\gamma} = (\gamma_1, \ldots, \gamma_K)^T$. If all patients had complete records, the parameters could be estimated by a penalized logistic likelihood, where $(\WH{\bs{\alpha}}, \WH{\bs{\beta}}, \WH{\bs{\gamma}}) = \argmin_{\bs{\alpha}, \bs{\beta}, \bs{\gamma}} \phi(\bs{\alpha}, \bs{\beta}, \bs{\gamma})$, and the objective function $\phi$ is defined by
\BEQ\label{eqn:objFn} 
	\phi(\bs{\alpha}, \bs{\beta}, \bs{\gamma}) = 
 	-{\rm loglik} (\bs{\alpha}, \bs{\beta}, \bs{\gamma}) + 
	P_{\lambda_1}(\bs{\beta}) + Q_{\lambda_2}(\bs{\gamma}).
\EEQ 
The  normalized   negative log-likelihood function is 
\BEA
- {\rm loglik} (\bs{\alpha}, \bs{\beta}, \bs{\gamma}) 
&=& 
	\frac{1}{N} \sum_{i=1}^K \sum_{j=1}^{n_i} \bigg[
	 \log\{1 + \exp({\bs Z}_{ij}^T \bs{\alpha}_1 + 
	 {\bs X}_i^T \bs{\alpha}_2 + \beta_i + \gamma_i)\} \NN  \\
&&	 \qquad	 - Y_{ij} ({\bs Z}_{ij}^T \bs{\alpha}_1 + 
	 {\bs X}_i^T \bs{\alpha}_2 + \beta_i + \gamma_i) \bigg]. \label{eqn:loglik}
\EEA
The second term $P_{\lambda_1}(\bs{\beta})$ is a fusion penalty that stems from the Lagrangian of \eqref{eqn:const1}, where 
$
P_{\lambda_1}(\bs{\beta}) = \lambda_1 \sum_{i_1 < i_2} \rho_{i_1,i_2} |\beta_{i_1} - \beta_{i_2}|.
$
We use $\rho_{i_1,i_2} = 1/ d(\bs{S}_{i_1}, \bs{S}_{i_2})$, where the $d(\bs{S}_{i_1}, \bs{S}_{i_2})$ denotes a distance between $\bs{S}_{i_1}$ and $\bs{S}_{i_2}$. Here, geodistance is used to define {$d(\cdot,\cdot)$} but other measures of similarity can be employed.  %Other forms are also possible. For example, \cite{Tibshirani2011a} utilized the fusion penalty in geospatial smoothing, where $\rho_{i_1, i_2} = 1$ if two locations share border and $\rho_{i_1, i_2} = 0$ otherwise. %such as the inverse distance, $\rho_{i_1, i_2} = 1 / \|\bs{S}_{i_1} - \bs{S}_{i_2}\|_2$.
Without loss of generality, we assume $\max_{i_1, i_2} \rho_{i_1, i_2} = 1$, otherwise we can normalize it by redefining $\rho_{i_1, i_2}$ with $\rho_{i_1, i_2} / \max_{i_1, i_2} \rho_{i_1, i_2}$. 
Since the computational cost of the optimization involving fusion penalty increases quadratically in the number of nonzero $\rho_{i_1,i_2}$'s, 
one may want to retain a few   $\rho_{i_1, i_2}$s with large values and truncate the others at zero for  ease of computation.

The third term, $Q_{\lambda_2}(\bs{\gamma}) =  \sum_{i=1}^K n_{i} q_{\lambda_2}(\gamma_i)/N$, is a  sparse penalty that is a relaxation of the Lagrangian of \eqref{eqn:const2}, where $q_{\lambda}(\cdot)$ is a univariate penalty function. In particular, we consider the hard penalty function as proposed in \cite{She2011},  
$
q_{\lambda} (t) = (\lambda |t| - t^2/2 ) I(t < \lambda) + {\lambda^2}/{2} I(t \geq \lambda). 
$
The hard penalty results in a nonconvex formulation on \eqref{eqn:objFn}, which guarantees convergence to a local minima. %Hence, the choice of the initial parameters matters.  
We weigh the $i$-th penalty in $Q_{\lambda_2}(\bs{\gamma}) $ by $n_i$ such that  subjects across different regions are penalized with the same amount.  

% Solving algorithm
\subsection{Optimization algorithm}\label{subsec:optim}

We developed an alternating minimization algorithm. It alternately  updates  $\bs{\alpha}$, $\bs{\beta}$, and $\bs{\gamma}$, each time minimizing one of them while keeping the others fixed.  Denote the current iterates by $\bs{\alpha}^{(t)}$, $\bs{\beta}^{(t)}$, and $\bs{\gamma}^{(t)}$. In addition, we denote   $\bs{Q}_{ij} = (\bs{Z}_{ij}^T, \bs{X}_i^T)$. Then $\bs{Q}_{ij}^T \bs{\alpha} = \bs{Z}_{ij}^T \bs{\alpha}_1 + {\bs X}_i^T \bs{\alpha}_2$.

\paragraph{\bf Updating $\bs{\alpha}$}  Fix $\bs{\beta} = \bs{\beta}^{(t)}$ and $\bs{\gamma} = \bs{\gamma}^{(t)}$. The objective function is equivalent to 
\[
\phi \left(\bs{\alpha}, \bs{\beta}^{(t)}, \bs{\gamma}^{(t)}\right) = \frac{1}{N} \sum_{i=1}^K \sum_{j=1}^{n_i} 
	\left[ \log\{1 + \exp \left(\bs{Q}_{ij}^T \bs{\alpha} + \mu_{ij}^{(t)} \right)\} 
	 - Y_{ij} \left(\bs{Q}_{ij}^T \bs{\alpha} + \mu_{ij}^{(t)} \right) \right]
\] 
with $\mu_{ij}^{(t)} = \beta_i^{(t)} + \gamma_i^{(t)}$, which corresponds to a classical logistic regression on $N$ individuals. One can run standard packages (such as \texttt{glm} in \texttt{R}) to obtain $\bs{\alpha}^{(t+1)}$. 
%One can run standard packages (such as \texttt{glm} in \texttt{R}) by specifying a covariate matrix as ${\bf Q}_{N \times q}$ which consists of $\bs{Q}_{ij}^T$ in each row, a response column as $\{Y_{ij}\}$ and an offset column as $\{\mu_{ij}^{(t)}\}$. %, and an individual-weight column as $\{w_{ij}\}$.  
%$\bs{\alpha}^{(t+1)}$ can be obtained subsequently.

\paragraph{\bf Updating $\bs{\beta}$} Fix $\bs{\alpha} = \bs{\alpha}^{(t+1)}$ and $\bs{\gamma} = \bs{\gamma}^{(t)}$, then
\begin{gather}
\phi \left(\bs{\alpha}^{(t+1)}, \bs{\beta}, \bs{\gamma}^{(t)}\right) 
=
	\underbrace{\frac{1}{N} \sum_{i=1}^K \sum_{j=1}^{n_i} \left[
	\log \left\{1 + \exp \left(\beta_i + \theta_{ij}^{(t)}\right)\right\} 
	- Y_{ij} \left(\beta_i + \theta_{ij}^{(t)} \right) \right]}_{=: ~l(\bs{\beta})} \NN \\
	\qquad + \, \lambda_1  \sum_{i_1 < i_2} \rho_{i_1,i_2} |\beta_{i_1} - \beta_{i_2}|, \NN
\end{gather}
where $\theta_{ij}^{(t)} = \bs{Q}_{ij}^T \bs{\alpha}^{(t+1)} + \gamma_i^{(t)}$ for each $i$ and $j$. 
For simplicity, define $\psi(\bs{\beta}) = \phi \left(\bs{\alpha}^{(t+1)}, \bs{\beta}, \bs{\gamma}^{(t)}\right)$, which is convex in $\bs{\beta}$. To update $\bs{\beta}^{(t)}$, we propose to minimize a surrogate objective function in which $l(\bs{\beta})$ is replaced by its local quadratic approximation  around $\bs{\beta}^{(t)}$. %The approximated problem objective function becomes This enables us to recast the problem as the least square with fusion penalties, which can be solved efficiently by off-the-shelf solvers.}
The same strategy was applied in implementing \texttt{R} package \texttt{glmnet} to iteratively descend the objective function of the generalized linear model with the elastic-net penalty \citep{Friedman2010a}. 

Write the second-order Taylor expansion of $l(\bs{\beta})$ at $\bs{\beta}^{(t)}$  as
\[
\TD{l}(\bs{\beta};\bs{\beta}^{(t)}) = l(\bs{\beta}^{(t)}) + \nabla_{\bs{\beta}} l(\bs{\beta}^{(t)})^T (\bs{\beta} - \bs{\beta}^{(t)}) + \frac{1}{2} (\bs{\beta} - \bs{\beta}^{(t)})^T \nabla_{\bs{\beta}\bs{\beta}}^2 l(\bs{\beta}^{(t)}) (\bs{\beta} - \bs{\beta}^{(t)}),
\]
where $\nabla_{\bs{\beta}}$ and $\nabla_{\bs{\beta}\bs{\beta}}^2$ are the first and the second derivative operators with respect to $\bs{\beta}$. Define the surrogate  objective function as
$
\TD{\psi}(\bs{\beta};\bs{\beta}^{(t)}) = \TD{l}(\bs{\beta};\bs{\beta}^{(t)}) + P_{\lambda_1}(\bs{\beta}).
$ 
We calculate $\tilde {\bs \beta}  = \argmin_{\bs{\beta}} \TD{\psi}(\bs{\beta}; \bs \beta^{(t)})$, where 
\[
\tilde {\bs \beta}  = \argmin_{\bs{\beta}} \left[ 
	\frac{1}{2} \sum_{i=1}^K  a_i^{(t)} \left(\beta_i - b_i^{(t)} \right)^2
	+ \lambda_1  \sum_{i_1 < i_2} \rho_{i_1,i_2} |\beta_{i_1} - \beta_{i_2}|
	\right],
\] 
with
\begin{gather}
a_i^{(t)} = \sum_{j=1}^{n_i} 
	\frac{ \exp\left( \beta_i^{(t)} + \theta_{ij}^{(t)} \right)}
	{ \left\{ 1 + \exp\left( \beta_i^{(t)} + \theta_{ij}^{(t)} \right) \right\}^2 };  \quad  
	 b_i^{(t)} = \beta_i^{(t)} - \frac{1}{a_i^{(t)}} \sum_{j=1}^{n_i} \left[  
	\frac{ \exp\left( \beta_i^{(t)} + \theta_{ij}^{(t)} \right)}
	{ 1 + \exp\left( \beta_i^{(t)} + \theta_{ij}^{(t)} \right) }
	- Y_{ij} \right]. \NN
\end{gather}
%This can be solved with packages implementing the generalized lasso penalty. 
 {For the calculation of $\tilde {\bs \beta}$, w}e applied the majorization-minimization algorithm proposed by \cite{Yu2015}, which yields a stable solution and can be easily implemented. 

 To ensure $\psi (\bs{\beta}^{(t)}) \geq \psi (\tilde{\bs{\beta}})$,  we adopt \cite{Lee2016d}'s one-step modification of $\tilde{\bs{\beta}}$: if $\psi (\bs{\beta}^{(t)}) \geq \psi (\TD{\bs{\beta}})$,  let $\bs{\beta}^{(t+1)} = \tilde{\bs{\beta}}$; otherwise,  $\bs{\beta}^{(t+1)} = \tilde h\tilde{\bs{\beta}} + (1-\tilde h)\bs{\beta}^{(t)},$
where  
$
\tilde{h}  = \argmin_{h \in [0,1]}  \psi\left(h\tilde{\bs{\beta}} + (1-h)\bs{\beta}^{(t)}\right).
$
%The optimization for \eqref{eqn:MLQA5} can be done simply by a grid search.
We will show in Proposition \ref{prop:1} that  $\tilde{h}$  always exists and  {$\psi (\bs{\beta}^{(t)}) \geq \psi (\bs{\beta}^{(t+1)})$} holds over iterations.

\paragraph{\bf Updating $\bs{\gamma}$} Given that $\bs{\alpha} = \bs{\alpha}^{(t+1)}$ and $\bs{\beta} = \bs{\beta}^{(t+1)}$,
 %the objective function can be simplified as 
\begin{gather}
\phi \left(\bs{\alpha}^{(t+1)}, \bs{\beta}^{(t+1)}, \bs{\gamma}\right) 
= \frac{1}{N}
	\sum_{i=1}^K \sum_{j=1}^{n_i} \left[
	\log \left\{1 + \exp\left(\gamma_i + \nu_{ij}^{(t)}\right) \right\} 
	- Y_{ij} \left(\gamma_i + \nu_{ij}^{(t)}\right) \right] \NN \\
 \qquad	+  \frac{1}{N} \sum_{i=1}^K n_i q_{\lambda_2}(\gamma_i), \NN
\end{gather}
where $\nu_{ij}^{(t)} = \bs{Q}_{ij}^T \bs{\alpha}^{(t+1)} + \beta_i^{(t+1)}$. With a slight abuse of notation, we define a univariate objective function $\phi_i (\gamma)$ and a loss function $l_i (\gamma)$ ($i = 1, \ldots, K$) as 
\[
\phi_i (\gamma) = \underbrace{\sum_{j=1}^{n_i} \left[
	\log \left\{1 + \exp\left(\gamma + \nu_{ij}^{(t)}\right) \right\} 
	- Y_{ij} \left(\gamma + \nu_{ij}^{(t)}\right) \right]}_{l_i (\gamma)} 
	+ \, n_i q_{\lambda_2}(\gamma).
\] 
Clearly $\phi \left(\bs{\alpha}^{(t+1)}, \bs{\beta}^{(t+1)}, \bs{\gamma}\right) = N^{-1} \sum_{i=1}^K \phi_i (\gamma_i)$. 
 Thus, it suffices to optimize $K$  univariate functions $\phi_i (\cdot)$, $i=1, \ldots, K$.
Although each $\phi_i (\gamma)$ is nonconvex, we can find a global optimum of $\phi_i$ as follows.
Let $\TD{t} = \argmin_{t \in \MBR} l_i (t)$. 
Since $q_{\lambda_2}(\cdot)$ is constant outside $[-\lambda_2, \lambda_2]$,
a minimizer of $\phi_i (\cdot)$ either lies on  $[-\lambda_2, \lambda_2]$ or equals to $\TD{t}$. 
Hence, we propose a grid search: let $\{t_1, \ldots, t_T\} \subseteq [-\lambda_2, \lambda_2]$ and put $\WH{\gamma}_i^{(t+1)} = \argmin_{\gamma \in \{\TD{t}, t_1, \ldots, t_T \}} \phi_i (t)$.

Details of the algorithm are provided in the Web Supplementary Materials. The following property is guaranteed by the proposed algorithm.
\begin{proposition}\label{prop:1} Assume that for each $i$, there exist $j_1, j_2$ such that $Y_{ij_1} = 0$ and $Y_{ij_2} = 1$.
For any choice of $\bs{\alpha}^{(t)}$, $\bs{\beta}^{(t)}$ and $\bs{\gamma}^{(t)}$, the updated iterates $\bs{\alpha}^{(t+1)}$, $\bs{\beta}^{(t+1)}$ and $\bs{\gamma}^{(t+1)}$ by Algorithm  1 in the Web Supplementary Materials   satisfy a monotone decreasing property:
$
\phi \left(\bs{\alpha}^{(t)}, \bs{\beta}^{(t)}, \bs{\gamma}^{(t)}\right) 
	\geq \phi \left(\bs{\alpha}^{(t+1)}, \bs{\beta}^{(t)}, \bs{\gamma}^{(t)}\right)
	\geq \phi \left(\bs{\alpha}^{(t+1)}, \bs{\beta}^{(t+1)}, \bs{\gamma}^{(t)}\right)
	\geq \phi \left(\bs{\alpha}^{(t+1)}, \bs{\beta}^{(t+1)}, \bs{\gamma}^{(t+1)}\right)
$.
\end{proposition}
The proof is deferred to the Web Supplementary Materials. The assumption indicates that the naive prevalence rate $\sum_{j=1}^{n_i} Y_{ij} / n_i$ lies on $(0,1)$ for each $i$, which is crucial to guarantee the existence of the optima at each step. 
By Proposition \ref{prop:1}, any limit point of $\{(\bs{\alpha}^{(t)}, \bs{\beta}^{(t)}, \bs{\gamma}^{(t)})\}$ is a stationary point  if $\phi$ is continuous. Since the objective function $\phi$ is nonconvex due to the nonconvexity of the hard penalty function, the proposed algorithm can only guarantee the convergence to a local optimum and requires a careful choice of the initial point. %If the objective function is optimized multiple times for a given set of tuning parameters, 
We could use the \emph{warm start strategy}, where the solution under the previous tuning parameter is used as the initial point for the next choice of tuning parameter. We applied this strategy in our simulation and data analysis, which showed a satisfactory performance. 

\noindent {\bf Remark.} Although the described algorithm handles a binomial likelihood function, it can be easily extended to other  (multilevel) generalized linear models. We can still solve $\bs{\alpha}$-step using an off-the-shelf package  (e.g. \texttt{glm} in \texttt{R}),  and $\bs{\beta}$- and $\bs{\gamma}$-steps  using the same strategies as outlined in the paper.

% Model selection (BIC)
\subsection{Choice of tuning parameter}

The proposed procedure involves the choice of $\lambda_1$ and $\lambda_2$. We implemented a model selection procedure to tune those parameters. Particularly, we used the modified Bayesian information criterion (BIC) proposed in \cite{She2011}, 
$
{\rm BIC}^*(\lambda_1, \lambda_2) = - 2 N \cdot {\rm loglik}(\WH{\bs{\alpha}}, \WH{\bs{\beta}}, \WH{\bs{\gamma}}) + {\rm DF} \cdot \left(1 + \log N \right).
$
Here, ${\rm loglik}(\WH{\bs{\alpha}}, \WH{\bs{\beta}}, \WH{\bs{\gamma}})$ is defined in  \eqref{eqn:loglik}, and
 the degrees of freedom (DF) is calculated by combining the DF calculated in the lasso and fused lasso regressions \citep{Tibshirani2005, Zou2007,Tibshirani2011a}, where
\BEA
{\rm DF} = ({\rm dimension~of~ }\widehat{\bs{\alpha}}) + ({\rm \#~of~distinct~values~of~}\widehat{\bs{\beta}})  +({\rm \#~of~nonzero~values~of~} \widehat{\bs{\gamma}}).\label{eqn:DF} 
\EEA
We searched  for the $(\lambda_1, \lambda_2)$  among a candidate set that minimizes the ${\rm BIC}^*(\lambda_1, \lambda_2)$. An alternative tuning method is  cross-validation,  which however, is much more computationally demanding. We use the  ${\rm BIC}^*(\lambda_1, \lambda_2)$ throughout.

\subsection{Weighting to account for missingness and selection bias}
As indicated in the previous sections, our dataset involves a large number of missing values for the obese indicators ($Y_{ij}$). Furthermore, the data may not be directly comparable to a national sample. For example,  in geographic areas and population groups that have traditionally experienced disparities in healthcare access and outcomes, the adoption of EHRs may not be as widespread.  These locations could be less represented. 
We consider  a two-step weighting procedure to adjust for both missing BMI values and selection bias.

The first step is to account for the missingness of BMI. We assumed the missing at random (MAR), where the probability of missing BMI is independent of its response conditional on the covariates  \citep{little2014statistical}.  Let $R_{ij} = 1$ if $Y_{ij}$ is observed and $R_{ij} =  0$ otherwise. The weight was defined as the inverse probability of observing BMI, $\MBP(R_{ij}=1 | \bs Z_{ij}, \bs X_i)$. This was unknown in practice. Hence we estimated it with a logistic regression using the observed data.  %The estimated weight  is $\widehat{\MBP}(R_{ij}=1 | \bs Z_{ij}, \bs X_i)^{-1}$.
The second step is to adjust for the population distribution of age, sex, and race/ethnicity. We applied a post-stratification correction  using the 2012 national census data. The final weight for each subject was the product of the inverse probability weight and the post-stratification weight. The objective function and subsequent procedures are modified subsequently. Details can be found in the Web Supplementary Materials. 
 
\section{Simulation studies}\label{sec:simul}

%\BIT
%\item {\bf Target of comparison...}: (if any..) an outlier discovery from linear (mixed) effect model?
%\EIT
%a, b, c, d, e, f ->
%a, X, b, c, d, e

We compared the proposed method with classic generalized mixed effect model (GLMM) and
%(b)  Bayesian conditional autoregressive model (BCAR, see  \cite{Lee2013a}); and
  the covariate-adjusted spatial scan statistic proposed by \cite{Jung2009} (Scan Statistic).  
The GLMM assumes ${\rm logit}(p_{ij}) = \bs{Z}_{ij}^T \bs{\alpha}_1 + \bs{X}_i \bs{\alpha}_2 + b_i + \delta$ where $b_i \sim \MC{N}(0, \sigma^2)$ {and $\delta$ is the global intercept}. %In BCAR, spatial autoregressive prior on $\{b_i\}_{k=1}^K$ is imposed to address regional dependency. 
To implement GLMM, we used the function \texttt{glmer()} of \texttt{R} package \texttt{lme4}. %We also conducted outlier detection based on GLMM, where the function  \texttt{ranef()} was used to obtain the predicted random effect. 
Let $\WH{b}_i$ be the predicted random effect of the $i$-th region from the fitted model. The $i$-th region was declared as an outlier if $|\WH{b}_i| > 2.5 \WH{\sigma}$.  
%We used the function \texttt{S.CARmultilevel()} of \texttt{R} package \texttt{CARBayes} to carry out BCAR. Outlier detection was conducted similarly as in GLMM; the only difference was that we considered the the posterior mean of $b_i$ as the predicted random effect. 
The scan statistic assumes ${\rm logit}(p_{ij}) = \bs{Z}_{ij}^T \bs{\alpha}_1 + \bs{X}_i \bs{\alpha}_2 + I(i \in S) \theta  + \delta$ and searches for  $S$, a cluster of regions, such that the null hypothesis of $H_{0}: \theta=0$ is rejected  {with the largest likelihood ratio statistic}. We also included three ``oracle'' versions of our methods, where $\phi$ is minimized with respect to one of $\bs{\alpha}$, $\bs{\beta}$, or $\bs{\gamma}$ while the other two are set to their true values:  with respect to $\bs{\alpha}$ (Oracle $\bs{\alpha}$); with respect to $\bs{\beta}$ (Oracle $\bs{\beta}$); and  with respect to $\bs{\gamma}$ (Oracle $\bs{\gamma}$).

%\red{ was implemented by ourselves.} \sout{The covariate-adjusted scan statistic method outputs clusters with the highest likelihood ratio as outlier regions, indicating these regions either had the highest or the lowest estimated incidence rates.}  %-(e) represent the fundamental limitations of our method when estimating $\bs{\alpha}$, $\bs{\beta}$ and $\bs{\gamma}$, respectively.

%a, b, c, d, e, f ->
%a, X, b, c, d, e
%Those covariate effects and area-level prevalences have been typical estimation targets in the literature of obesity as well as other public health concerns.    

%\blue{\small (Reviewer 2 Minor comment: Please elaborate a bit more on the condition for a region to be treated as an outlier.)}

%\red{We generated simulated datasets of \eqref{eqn:model} as follows.} 
We considered $K$ $(K=20, 40)$ regions where the number of subjects in each region was $n$ $(n=50, 100)$. %to see if our method shows a consistency behavior. 
We generated  $\bs{Q}_{ij} = (Z_{ij}, X_i)^T$, where $Z_{ij}$ and $X_i$ were drawn from Bernoulli distributions with a probability of 0.5. We set $\bs{\alpha} = (\alpha_1, \alpha_2) = (-0.2, 0.2)$.  For simplicity, we simulated $K$ locations on a one-dimensional line with $S_i \sim {\rm Unif}(5, 95)$, $i=1,\ldots,K$. %Hence, the distance between any two locations $S_{i_1}$ and $S_{i_2}$, and subsequently $\rho_{i_1, i_2}$, can be easily calculated. 
$\beta_i$ was set to ${\rm logit}(0.4)$ if $5 \leq S_i < 35$, ${\rm logit}(0.5)$ if $35 \leq S_i < 65$, and ${\rm logit}(0.6)$ if $65 \leq S_i \leq 95$. We randomly chose $K_O$ regions, where $\gamma_i = 2$ for $\lfloor K_0 / 2 \rfloor$ regions  ($\lfloor t \rfloor$ is the maximum integer no larger than $t$) and $\gamma_i = -2$ for the remaining.  Thus, those regions with  $\gamma_i \neq 0$ should be regarded as the true outlier regions. 
 We varied the number of outliers so that $K_O/K = 0\%, 5\%, 10\%, 15\%$. 
 For each scenario, we repeatedly generated   1000 datasets. We applied different methods on each dataset and evaluated the performance metrics. The performance measures were averaged over the 1000 replications. The tuning parameters were selected using the proposed modified BIC, among a pre-defined candidate set with $\lambda_1 \in [2^{-2}, 2^{12}]$ and $\lambda_2 \in [2^{-5}, 2^{2}]$.

We compared the proposed method with GLMM and Oracle $\bs{\alpha}$ in terms of the bias of the individual-level covariate effect, $\WH{\alpha}_1 - \alpha_1$, community-level covariate effect, $\WH{\alpha}_2 - \alpha_2$, and the root mean squared error (RMSE) of the region-level prevalence rates, $\sqrt{\sum_{i=1}^{K}(\WH{p}_i - p_i)^2 / K}$. Here, $p_i = \MBE(Y_{ij}|\bs{X}_i)$ and $\WH{p}_i$ is the empirical average of the estimated individual-level prevalence estimates, $\WH{p}_{ij}$, taken over $j=1,\ldots, n_i$. The results are presented in Table \ref{tbl:trad}. The biases of estimating $\alpha_1$, the individual-level covariate effect, were close to zero in the proposed method, especially when both $n$ and $K$ increase. The confidence intervals from the proposed and GLMM were comparable. %, while those from the BCAR were slightly wider. 
The biases of $\WH{\alpha}_2$, the region-level covariate effect,  were reduced as $K$ increased in the proposed.  The performances of the proposed and the method Oracle $\bs{\alpha}$ were similar in terms of the biases. In addition, they were barely affected by the increased proportion of outliers. % \sout{On the other hand, the biases of $\WH\alpha_2$ in the GLMM increased with a larger proportion of outliers.}
On the other hand, $\WH\alpha_2$ in the GLMM %and BCAR 
had increased variability with a larger proportion of outliers. 
 The RMSEs of $\WH{p}_i$ were smaller in the proposed than in the GLMM,
 %and BCAR 
 although as anticipated,  larger than the case where only $\bs{\alpha}$ was unknown.

\begin{table}[t]
\caption{95\% confidence intervals of biases of $\WH{\alpha}_1$ and $\WH{\alpha}_2$, and RMSE of $\{\WH{p}_i\}_{i=1}^K$, over 1000 replications.}
\label{tbl:trad}
\begin{center}
{\scriptsize
\begin{tabular}{c|l|ccc|ccc}
\hline
 &  & \multicolumn{3}{c|}{$n=50$ per region} &  \multicolumn{3}{c}{$n=100$ per region} \\ \cline{3-8}
$K_O/K$ & 	Method	&  Bias  & Bias & RMSE & Bias & Bias & RMSE \\ 
 &  & of $\WH{\alpha}_1$ &  of $\WH{\alpha}_2$ &  of $\{\WH{p}_{i}\}$ & of $\WH{\alpha}_1$ &  of $\WH{\alpha}_2$ &  of $\{\WH{p}_{i}\}$ \\\hline
  \multicolumn{8}{c}{$K=20$ regions}  \\ \hline
    & Proposed &  -.004 (-.012, .004) &  -.002 (-.014, .009) &  .053 &  .002 (-.004, .008) &  -.016 (-.025, -.007) &  .040 \\ 
 0\% &  GLMM &  -.004 (-.012, .004) &  -.015 (-.027, -.002) &  .057 &  .002 (-.004, .007) &  -.019 (-.030, -.008) &  .044 \\ 
    &  Oracle $\bs{\alpha}$&  -.002 (-.009, .005) &  .002 (-.004, .008) &  .016 &  .002 (-.003, .007) &  -.004 (-.009, .000) &  .011 \\  \hline
    & Proposed &  -.003 (-.011, .006) &  -.003 (-.015, .009) &  .053 &  .002 (-.004, .008) &  -.016 (-.025, -.006) &  .040 \\ 
 5\% &  GLMM &  -.003 (-.011, .005) &  -.016 (-.033, .001) &  .062 &  .001 (-.005, .007) &  -.020 (-.036, -.004) &  .046 \\ 
    &  Oracle $\bs{\alpha}$&  -.001 (-.008, .006) &  .002 (-.004, .009) &  .016 &  .002 (-.002, .007) &  -.003 (-.008, .002) &  .011 \\  \hline
    & Proposed &  -.003 (-.012, .005) &  -.004 (-.016, .009) &  .055 &  .002 (-.003, .008) &  -.017 (-.027, -.007) &  .040 \\ 
 10\% &  GLMM &  -.004 (-.013, .004) &  -.001 (-.021, .019) &  .063 &  .002 (-.004, .007) &  -.006 (-.026, .014) &  .046 \\ 
    &  Oracle $\bs{\alpha}$&  -.002 (-.009, .005) &  .003 (-.003, .010) &  .016 &  .003 (-.002, .008) &  -.003 (-.008, .002) &  .011 \\  \hline
    & Proposed &  -.004 (-.012, .005) &  -.002 (-.016, .011) &  .055 &  .002 (-.004, .008) &  -.016 (-.026, -.006) &  .041 \\ 
 15\% &  GLMM &  -.005 (-.013, .003) &  .009 (-.014, .033) &  .063 &  .001 (-.005, .007) &  .002 (-.021, .026) &  .046 \\ 
    &  Oracle $\bs{\alpha}$&  -.003 (-.010, .004) &  .005 (-.002, .011) &  .016 &  .002 (-.003, .007) &  -.004 (-.009, .001) &  .011 \\  \hline
\multicolumn{8}{c}{$K=40$ regions}  \\ \hline 
    & Proposed &  -.002 (-.008, .004) &  .002 (-.005, .009) &  .043 &  -.000 (-.004, .004) &  -.003 (-.008, .003) &  .033 \\ 
 0\% &  GLMM &  -.002 (-.008, .003) &  .001 (-.008, .009) &  .055 &  -.001 (-.005, .003) &  -.005 (-.012, .003) &  .043 \\ 
    &  Oracle $\bs{\alpha}$&  -.002 (-.007, .002) &  .003 (-.001, .008) &  .012 &  .001 (-.002, .004) &  -.001 (-.004, .002) &  .008 \\  \hline
    & Proposed &  -.003 (-.009, .003) &  .004 (-.004, .011) &  .044 &  .000 (-.004, .004) &  -.002 (-.008, .003) &  .034 \\ 
 5\% &  GLMM &  -.004 (-.010, .002) &  .003 (-.008, .014) &  .060 &  -.000 (-.005, .004) &  -.003 (-.014, .008) &  .045 \\ 
    &  Oracle $\bs{\alpha}$&  -.002 (-.007, .002) &  .004 (-.001, .009) &  .012 &  .001 (-.002, .005) &  -.001 (-.005, .002) &  .008 \\  \hline
    & Proposed &  -.004 (-.010, .002) &  .003 (-.004, .011) &  .045 &  -.001 (-.005, .003) &  -.001 (-.006, .005) &  .034 \\ 
 10\% &  GLMM &  -.004 (-.010, .002) &  -.003 (-.017, .011) &  .062 &  -.002 (-.006, .002) &  -.007 (-.021, .007) &  .046 \\ 
    &  Oracle $\bs{\alpha}$&  -.003 (-.008, .001) &  .004 (-.001, .008) &  .012 &  .001 (-.003, .004) &  .000 (-.003, .004) &  .008 \\  \hline
    & Proposed &  -.004 (-.010, .002) &  .005 (-.003, .013) &  .046 &  -.000 (-.005, .004) &  -.001 (-.007, .004) &  .034 \\ 
 15\% &  GLMM &  -.005 (-.011, .001) &  -.002 (-.018, .014) &  .063 &  -.001 (-.005, .003) &  -.006 (-.022, .010) &  .046 \\ 
    &  Oracle $\bs{\alpha}$&  -.003 (-.008, .001) &  .004 (-.001, .009) &  .012 &  .001 (-.002, .005) &  .000 (-.003, .004) &  .008 \\  \hline 
\end{tabular}
}
\end{center}
\end{table}

\begin{figure}[t!]
\caption{ RMSE of $\WH{\bs{\beta}}$, varying the number of outliers over 1000 replications. }
\label{fig:RMSE}
\centering{
\includegraphics[scale=0.8]{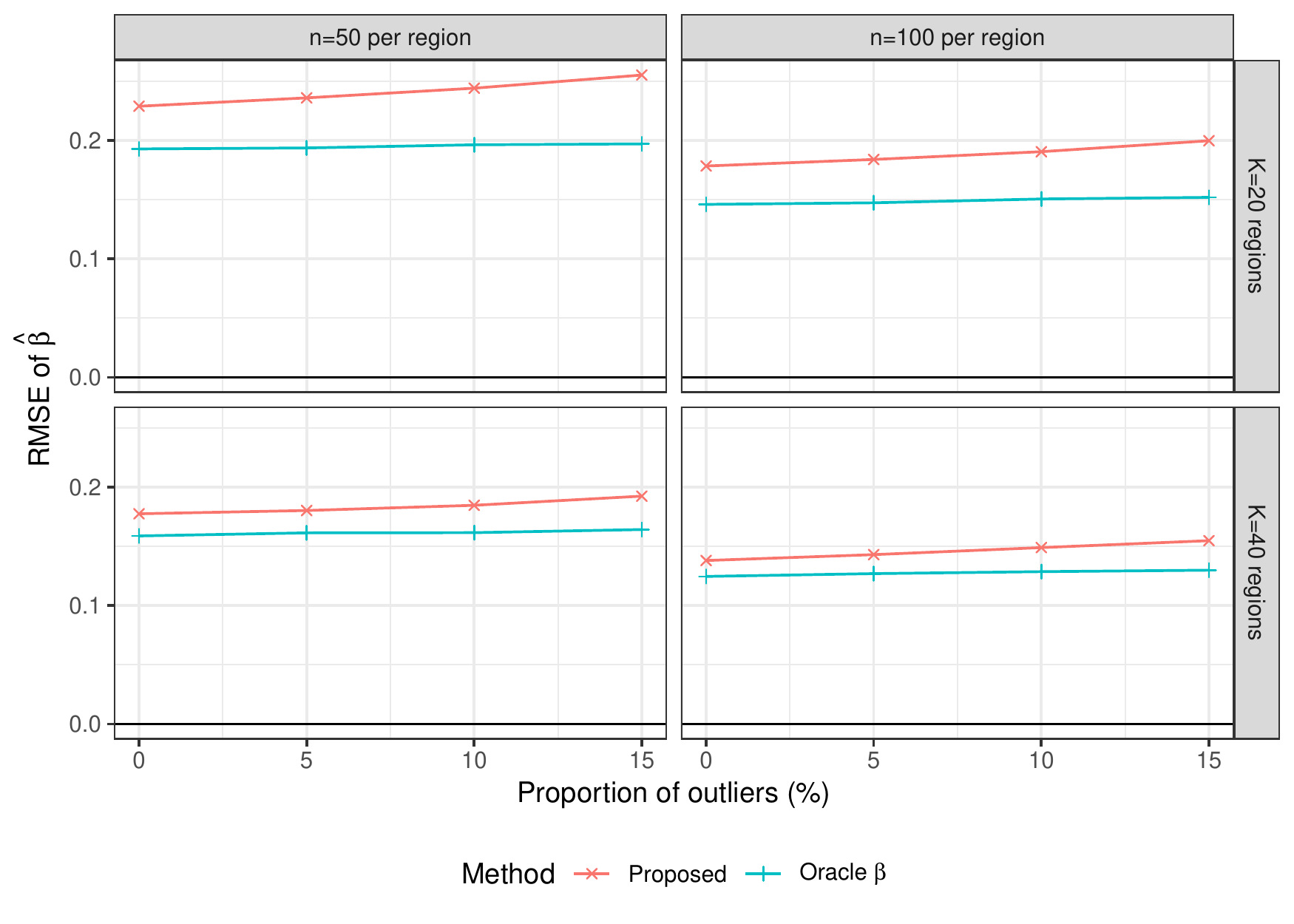}
}
\end{figure}

We further compared the RMSE of $\bs{\beta}$, $\sqrt{\sum_{i=1}^K (\WH{\beta}_i - \beta_i)^2/K}$, from the proposed method and Oracle $\bs\beta$. {Figure \ref{fig:RMSE}} shows that the RMSE of $\WH{\bs{\beta}}$ decreased when $n$ or $K$ increased, and slightly increased with a larger proportion of outliers. Compared with Oracle $\bs{\beta}$, the difference between two methods became smaller with an increasing $K$. This indicates that the proposed method provides a good estimate of the baseline obesity rate.

\begin{figure}[t!]
\caption{ MCC, varying the number of outliers over 1000 replications. }
\label{fig:MCC}
\centering{
\includegraphics[scale=0.8]{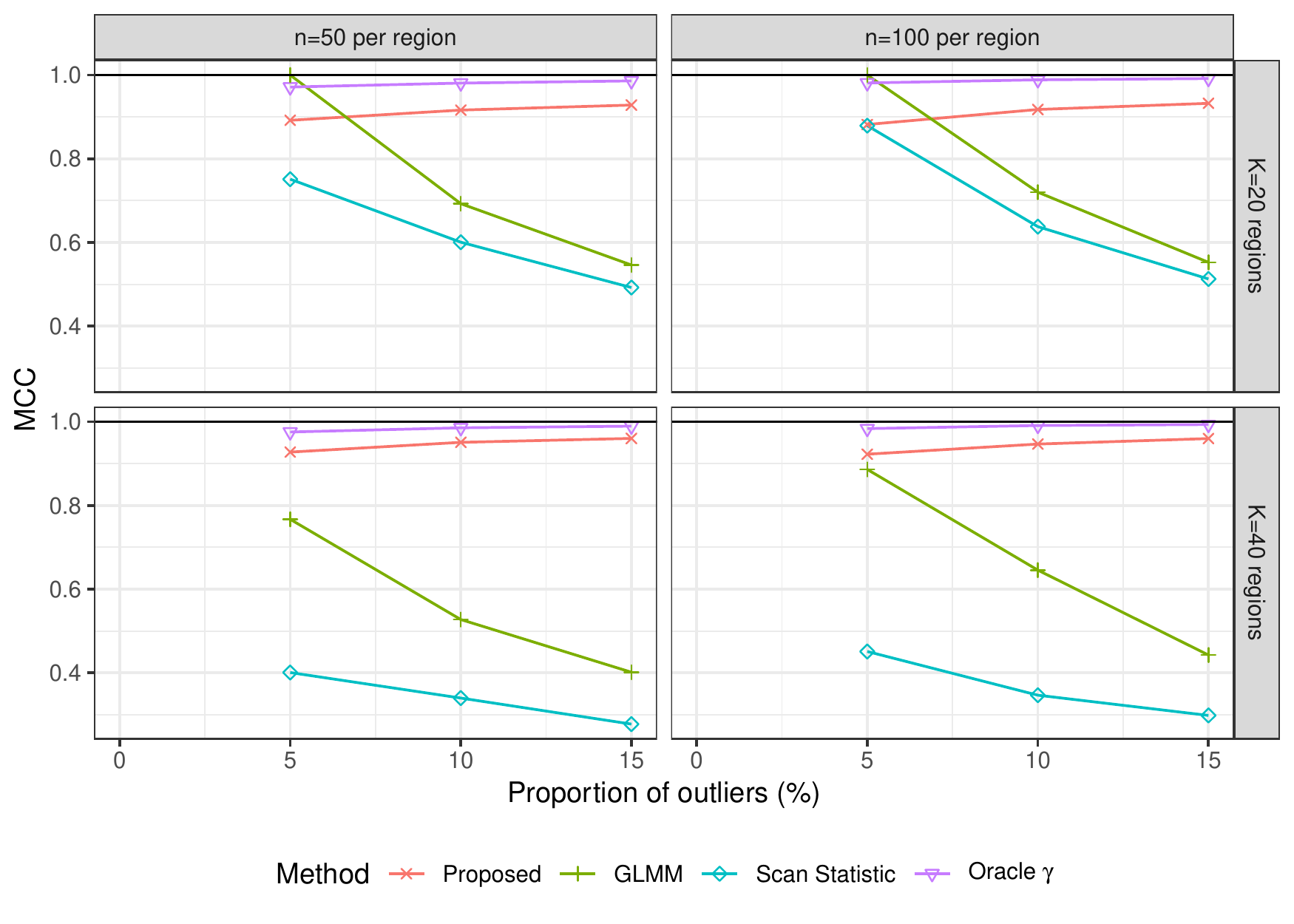}
}
\end{figure}

Finally, to compare the performances in outlier detection, we used Matthews Correlation Coefficient (MCC), defined by
\[
{\rm MCC = \frac{TP \cdot TN - FP \cdot FN}{\sqrt{(TP+FP)(TP+FN)(TN+FP)(TN+FN)}} }. 
\]
Here, TP stands for true positive, where the detected outlier region is indeed an outlier; TN stands for true negative, where the labeled normal region is normal;  FP stands for false positive, where the detected outlier region is actually normal; and FN stands for false negative, where the labelled normal region is actually an outlier.  A higher value of MCC is preferred, where ${\rm MCC} =1$ indicates a perfect classifier and ${\rm MCC} =0$ indicates a random guess. We evaluated the MCC on the proposed, GLMM, Scan Statistic, and Oracle $\bs\gamma$.  The MCCs are presented in Figure \ref{fig:MCC}. 
The MCC of the Scan Statistic was around zero, even when $n$ and $K$ were increased, indicating that the Scan Statistic failed to detect multiple outlier regions. %\blue{(YZ: did you update this paragraph? I think now scan statistic incorporates individual covariates?)} \red{(YC: I updated and did not include the covariate effects of the scan statistic. The fitted covariate effects of the scan statistic become different according to which cluster $S$ is selected and tested. Although we identified such one $S$ with the highest $p$-value (one cluster), the scan stat procedure in general picks several clusters in the order of lowest $p$-values. Thus, it might look weird if we pick one $S$ and report corresponding coefficients.  \cite{Jung2009} does not reported the covariate effects as well.)}
The MCC  of the GLMM  %and BCAR  
decreased when either the proportion of outlier  or the $K$ increased. This is consistent with the literature suggesting that residual-based outlier detection may not operate well with multiple outliers. The MCC of the proposed method was comparable to its oracle counterpart, Oracle $\bs{\beta}$. It  improved over increasing $K$ and stabilized over increasing proportion of outliers.  In summary, our method showed  promising performance in identifying outliers, especially when the proportion of outliers was increased.

\section{Application to the PHINEX database}\label{sec:application} 
 
We considered census  blockgroup  as the geographic unit, and  we excluded certain blockgroups with small sample sizes, following the guidelines of Behavioral Risk Factor Surveillance System \citep{CDC2016}. %The guided exclusion criteria were (1) the number of subjects within the blockgroup was less than 50, $n_i < 50$, or (2) the relative standard error to the mean was more than 0.3, i.e., $se(\WH{p}_i) / \WH{p}_i =  \sqrt{{\WH{p}_i(1-\WH{p}_i)}/{n_i}} / \WH{p}_i > 0.3$, where $\WH{p}_i = \sum_{j=1}^{n_i} Y_{ij}/ n_i$.  
% covariate choice
The individual covariates  $\bs{Z}_{ij}$ included sex, age as of 2012, race/ethnicity and  insurance status, %of the ($ij$)-th subject, 
and the region-level covariates $\bs{X}_i$ included urbanicity and EHI. Age was  categorized into 3 groups: 2--4 years, 5--9 years, and 10--14 years.
Race and ethnicity were combined into a single covariate, and categorized into 4 groups: Hispanic, non-Hispanic white, non-Hispanic black, and non-Hispanic other.
Patients with health service payor as commercial or Medicaid were included,  where a few subjects with no insurance were excluded. Urbanicity, ranging from 1 to 11, was categorized into 3 groups: urban (1-4), suburban (5-8), or rural (9-11). %, following \cite{ESRI2012} guidelines. 
We standardized EHI for numerical stability of the proposed algorithm. 

% distance  
The location $\bs{S}_i$ was defined by a vector of the longitude and latitude of the centroid of the $i$-th block group.  We constructed $\rho_{i_1, i_2}$ as the inverse of geodesic distances,  $\rho_{i_1, i_2} = 1 / d^{{\rm geo}}(\bs{S}_{i_1}, \bs{S}_{i_2})$, where $d^{{\rm geo}}(\bs{S}_{i_1}, \bs{S}_{i_2})$ denotes the greater circle distance between $\bs{S}_{i_1}$ and $\bs{S}_{i_2}$. 
For  the $i_1$-th region, we retained the $L$ largest $\rho_{i_1, i_2}$s and truncated the others at zero, where we treated $L$ as a tuning parameter.
A grid search on $\lambda_1 \in [2^{-1}, 2^{17}]$, $\lambda_2 \in [2^{-5}, 2^2]$ and $L \in \{3, 5, 7\}$ was conducted to find the best combination of tuning parameters.  %The selected tuning parameters chosen by the modified BIC were $\lambda_1^* = 2048$ and $\lambda_2^* = 0.463$. 
The confidence interval of each parameter was constructed using bootstrap over 1000 replications.
  
\begin{table}[t]
\centering
\caption{Fitted coefficients and  {confidence intervals (in parentheses)} for covariate effects.}
\bigskip
\label{tbl:alpha}
{
\begin{tabular}{l|cc}
\hline
 & \multicolumn{2}{c}{Model} \\
 & Proposed & GLMM  \\ \hline
\emph{Individual-level covariates} & \\
~~~~~ Sex (Base: Female)  &   \\
~~~~~$\sim$ Male &  .235 (.144, .305) &  .226 (.180, .271)\\
~~~~~Age at 2012 (Base: Pre-school) &   \\
~~~~~$\sim$ School-aged &  .568 (.429, .657) &  .562 (.491, .632)\\
~~~~~$\sim$ Adolescent &  .875 (.740, .967) &  .869 (.802, .936)\\
~~~~~Race/Ethnicity (Base: White, non-Hispanic) &  &  \\
~~~~~$\sim$ Black, non-Hispanic &  .437 (.313, .559) &  .434 (.360, .508)\\
~~~~~$\sim$ Other, non-Hispanic & .035 (-.133, .180) &  .042 (-.068, .153)\\
~~~~~$\sim$ Hispanic &  .680 (.569, .787) &  .667 (.606, .728)\\
~~~~~Insurance status (Base: Commercial) &   \\
~~~~~$\sim$ Medicaid & .522 (.428, .636) &  .509 (.453, .565)\\
\emph{Community-level covariates} & \\
~~~~~Urbanicity (Base: Urban) &   \\
~~~~~$\sim$ Suburban &  -.134  {(-.305, -.034)} &  .037 (-.078, .152)\\
~~~~~$\sim$ Rural & .125   {(-.163, .302)} &  .237 (.070, .405)\\
~~~~~Economic  Hardship Index (standardized) &  .120 {(-.001, .147)}  &  .143 (.095, .191)\\ \hline
\end{tabular}
}
\end{table}

\begin{figure}[t!]
\caption{Estimated baseline prevalence rates and the identified outliers in childhood obesity surveillance. Each   polygon represents a census  blockgroup. Top-left: Result from the proposed method. Outliers are marked as black (yellow) as above the trend, $\WH{\gamma}_i > 0$ (below the trend, $\WH{\gamma}_i < 0$). Top-right: Result from the GLMM. Outliers are marked as black (yellow) as above the trend, $\WH{b}_i  > 2.5 \WH{\sigma}$ (below the trend, $\WH{b}_i < 2.5 \WH{\sigma}$). Bottom-left: Discovered cluster by the Scan Statistic with the highest likelihood ratio, which was  {below} the trend.}
\label{fig:chol}
\centering{
\includegraphics[scale=0.62]{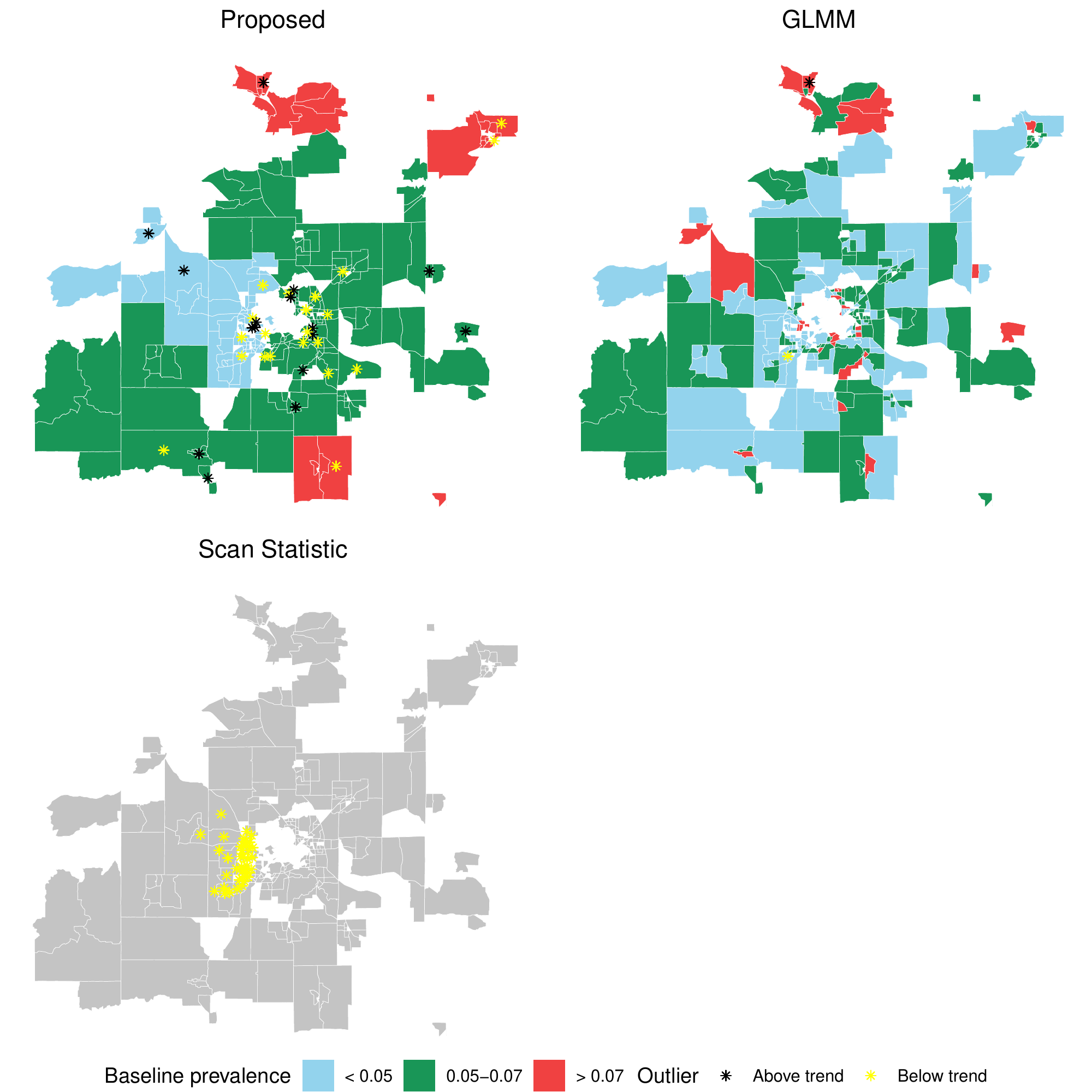}
}
\end{figure}

%indiv-level coefficients (alpha)
The estimated  $\WH{\bs{\alpha}}$s are summarized in Table \ref{tbl:alpha}. Overall, the proposed method had wider confidence intervals than the GLMM. This result was anticipated, given that our method had more parameters to estimate, which could lead to higher variabilities. The estimated coefficients from our model and the GLMM were comparable in general, except for the suburban effect. The obesity rate in females was lower compared to males, and younger children had lower obesity rates. Obesity rates in both non-Hispanic white and non-Hispanic other were lower than those in non-Hispanic black and Hispanic patients. 
The obesity prevalence was higher in subjects with Medicaid compared to those with commercial insurance. The EHI was positively associated with the estimated obesity rate.

%%% clusters
The fitted  baseline obesity rates are displayed in Figure \ref{fig:chol}, which appear to coincide with empirical knowledge of the greater Madison area.  %, capturing effects that are not fully explained by the covariates in the model. 
The lowest prevalence areas included the western portion of the Madison, Middleton, and Verona areas. It is known that these areas are recently developed and expanded, and include people who are generally younger and more socioeconomically advantaged compared to the surrounding areas. The intermediate prevalence areas, comprising the greater central and eastern Madison region, are more established, historic areas of the region, and are  known
to contain more stereotypical middle-class citizens.  The highest prevalence areas are clearly the most geographically distant from the center of Madison, and are also all outside of Dane county, which contains Madison. 
% END: Derek additions
%Although such socioeconomic conditions were included in our covariates, they might also have affected other endogeneous environmental factors not included in the covariates (e.g. school-level nutrition and retail food environment), which in turn may have been captured in the baseline obesity rates.

\begin{table}[t!]
\centering
\caption{The anonymized IDs of the outlier blockgroups, their sample sizes, crude obesity rates, fitted baseline obesity rates, adjusted obesity rates,  {and frequencies of detections over $B=1000$ bootstrap replications}. } 
\label{tbl:gamma}
{\footnotesize
\begin{tabular}{c|cc|cc|c}
\hline
blockgroup   & Unweighted & Crude & Fitted & Fitted & Frequency \\
 ID & sample & obesity rate  & baseline & adjusted & of  \\
 & size   &   &  obesity rate &  obesity rate & detection \\
 &    ($n_i$)    & 	($\WH{p}_i^{\rm crude}$) & ($\WH{p}_i^{\rm bsl}$) &   ($\WH{p}_i^{\rm adj}$) & ($\sum_i I(\WH{\gamma}_i \neq 0)/B$) \\  
  \hline      
  \multicolumn{6}{c}{Above the trend}\\
  \hline
7 & 60 & .432 & .097 (.061, .150) & .264 (.153, .351) &  .701 \\
23 & 91 & .234 & .048 (.042, .057) & .129 (.110, .167) &  .483 \\
24 & 93 & .206 & .048 (.042, .057) & .113 (.095, .128) &  .515 \\
25 & 104 & .207 & .048 (.042, .057) & .115 (.094, .139) &  .496 \\
83 & 96 & .291 & .058 (.047, .064) & .174 (.133, .179) &  .689 \\
85 & 91 & .356 & .058 (.047, .064) & .187 (.139, .196) &  .895 \\
100 & 62 & .288 & .058 (.047, .064) & .149 (.117, .155) &  .763 \\
102 & 53 & .335 & .058 (.047, .064) & .217 (.147, .246) &  .579 \\
124 & 66 & .207 & .058 (.047, .064) & .117 (.090, .136) &  .535 \\
200 & 93 & .218 & .058 (.047, .064) & .133 (.102, .151) &  .474 \\
212 & 100 & .180 & .048 (.042, .057) & .085 (.076, .094) &  .634 \\
244 & 71 & .203 & .053 (.044, .064) & .121 (.096, .135) &  .446 \\
245 & 94 & .248 & .053 (.044, .064) & .148 (.105, .184) &  .573 \\
252 & 68 & .278 & .056 (.046, .073) & .148 (.108, .179) &  .706 \\
254 & 82 & .204 & .058 (.048, .071) & .103 (.080, .111) &  .698 \\
264 & 74 & .257 & .048 (.042, .057) & .130 (.092, .157) &  .726 \\
\hline
  \multicolumn{6}{c}{Below the trend} \\
  \hline
22 & 110 & .063 & .048 (.042, .057) & .123 (.109, .151) & .363 \\
32 & 221 & .045 & .048 (.042, .057) & .092 (.080, .101) & .079 \\
35 & 146 & .067 & .048 (.042, .057) & .126 (.109, .151) & .300 \\
70 & 67 & .168 & .058 (.047, .064) & .283 (.203, .292) & .269 \\
73 & 109 & .175 & .058 (.047, .064) & .273 (.202, .279) & .136 \\
82 & 259 & .173 & .058 (.047, .064) & .329 (.209, .342) & .106 \\
94 & 134 & .105 & .058 (.047, .064) & .198 (.159, .210) & .383 \\
115 & 68 & .061 & .058 (.047, .064) & .157 (.120, .161) & .468 \\
118 & 192 & .141 & .058 (.047, .064) & .248 (.155, .253) & .062 \\
125 & 81 & .047 & .058 (.047, .064) & .115 (.091, .115) & .277 \\
127 & 295 & .051 & .058 (.047, .064) & .114 (.094, .121) & .195 \\
128 & 125 & .054 & .058 (.047, .064) & .154 (.122, .175) & .698 \\
136 & 466 & .038 & .048 (.042, .057) & .114 (.100, .132) & .684 \\
153 & 469 & .045 & .048 (.042, .057) & .095 (.085, .103) & .047 \\
159 & 202 & .081 & .058 (.047, .064) & .158 (.130, .176) & .345 \\
168 & 139 & .117 & .056 (.046, .071) & .221 (.169, .247) & .418 \\
186 & 66 & .097 & .058 (.047, .064) & .167 (.118, .184) & .204 \\
229 & 133 & .139 & .077 (.048, .127) & .299 (.189, .395) & .599 \\
235 & 61 & .139 & .077 (.048, .127) & .235 (.139, .312) & .365 \\
246 & 210 & .060 & .053 (.044, .064) & .138 (.106, .158) & .283 \\
262 & 90 & .076 & .081 (.050, .114) & .189 (.116, .237) & .459 \\
\hline
\end{tabular}
}
\end{table}

The proposed method identified several outliers. Aberrant locations with obesity rates above the trend ($\WH{\gamma}_i > 0$) and below the trend ($\WH{\gamma}_i < 0$) were shown as black and yellow, respectively, in Figure \ref{fig:chol}. We identified 6\% of blockgroups as outliers above the trend, and 8\% as below the trend. Results are presented in Table \ref{tbl:gamma}, including:  
\BIT
\item crude obesity rates,  $\WH{p}_i^{\rm crude} = \frac{1}{\sum_j w_{ij}I(R_{ij}=1)} \sum_j w_{ij} I(R_{ij}=1) Y_{ij}$;
\item baseline obesity rates, $\WH{p}_i^{\rm bsl} = {\rm logit}^{-1}(\WH{\beta}_i)$;
\item  obesity rates adjusted for covariates and outliers, 
$$
 \WH{p}_i^{\rm adj}  = \WH{\MBE}_ {\gamma_i = 0}(Y_{ij} | \bs{X}_i) = \frac{1}{\sum_j w_{ij}I(R_{ij}=1)} \sum_{j=1}^{n_i} w_{ij}I(R_{ij}=1) \cdot {\rm logit}^{-1} \left(
	\bs{Z}_{ij}^T \WH{\bs{\alpha}}_1 + \bs{X}_{i}^T \WH{\bs{\alpha}}_2 + \WH{\beta}_i \right),
$$
\EIT 
and frequencies of detections over $B=1000$ bootstrap replications. We note that the outlier identification is \emph{relative} to the fitted trend. For example, blockgroup  212 had an ordinary level of the estimated crude obesity rate (0.180). However, the crude rate was much higher than the fitted value of expected obesity prevalence (0.085). There existed unexplained information that could contribute to the elevated rate. Hence, it was declared as an outlier above the trend. The frequency of detection based on the bootstrap provides a glimpse of the uncertainty of aberrance. The outlier regions above the trend tended to have higher frequencies than those below the trend. 

The localized outbreak from our model may enable comparative investigations at granular level. For example, what  potential factors explain the outlier blockgroups that are significantly above or below the trend? 
Obesity prevalence is determined by the interplay of patient demographic characteristics, behaviors, and community environmental factors. Our model has accounted for only a subset of them. % At the individual level, it is sex, age, race/ethnicity, insurance status; and at the community level, it is urbanicity and EHI. 
Thus outliers could represent communities with meaningfully different environments than expected (e.g. much better or worse than average access to grocery stores, parks, and recreational facilities), and/or it could represent community members with behaviors that are substantially different than expected (e.g. much greater or less physical activity, substantially better or worse dietary habits, etc.).  
Based on our results, healthcare professionals could look into the risk factors within the outlier and compare these factors with its adjacent blockgroups.

We also applied the GLMM and the Scan Statistic for comparison.  
 As shown in Figure \ref{fig:chol}, the Scan Statistic identified midwestern Madison area as abnormal below trend, which appears to coincide with the lowest prevalence area identified by our method. However, it failed to inform outbreaks in obesity rates. These outbreaks could be captured by $\WH{\bs{\gamma}}$ in our model.  %\blue{YZ: Now we are using covariate adjusted scan stat, is this statement fair?} \red{(YC: the figure should be changed (current version is an unadjusted one). I plan to update the statement and figure when BCAR finishes running.)} 
The GLMM does not smooth the obesity pattern over the state, which might be difficult to explain and investigate. Thus, the proposed method could provide more informative and interpretable results.

\section{Concluding remarks}\label{sec:conclusion}

% {\small  \blue{(reviewer 2 major comment: how do you check that this assumption is correct?)}} \textcolor{red}{How did they assess this assumption in \citep{Knorr-Held2000} and  \citep{Forbes2013}?} 

% \blue{(comment: what are the reasons why you do not consider the whole data set?)}  \textcolor{red}{discuss generalization to longitudinal data.}

Motivated by  childhood obesity surveillance using routinely collected EHR data, we developed a multilevel penalized logistic regression model. We incorporated the fusion and the nonconvex sparsity penalty in the likelihood function, which enabled us to conduct regional smoothing and outlier detection simultaneously. While in this paper we only considered spatial surveillance, we are interested in generalizing the method to longitudinal data setup that conducts spatiotemporal surveillance. 

In our paper, we assume that BMI is MAR, which is not testable in observational data. The feasibility of MAR for EHRs data is an active area of research in recent years \citep{Snyder2018}. In the future, we can develop sensitivity analysis techniques that investigate sensitivity of the results to uncontrolled confounding \citep{Greenland2004}. 
  
%The proposed method showed superior performance over traditional methods. We applied the proposed method to the PHINEX database, which provided insights on childhood obesity at granular level. Methodologically, we extended a simultaneous fused lasso-based smoothing and outlier detection procedure to a generalized linear model setup. In particular, we utilized a nonconvex penalty for outlier detection.  Furthermore, the proposed alternating minimization algorithm could be generalized to handle a general convex loss in a straightforward fashion.

 Another future direction is to develop principled inferential procedures for the proposed work.  We could potentially use ideas from significance testing and confidence regions for  penalized procedures. For example, \cite{Hyun2016} developed a post-selection inference for generalized lasso applied to linear models.  A general selective inference procedure for penalized likelihood has been developed for $\ell_1$ penalty \citep{Taylor2016}. %\sout{Bootstrap could also work for inference after tuning parameter selection}  \citep{Efron2014}\sout{, although further exploration is needed for penalties other than the $\ell_1$ penalty.}

\section*{Acknowledgement}

This work is supported by  R21HD086754 awarded by the National Institutes of Health. It is also supported in part by NIH grant P30CA015704 and S10OD020069.
The authors thank Donghyeon Yu (Inha University, Incheon, South Korea) for sharing R codes on the majorization-minimization algorithm, and Albert Y. Kim (Amherst College, Amherst, MA) for sharing the source codes for the \texttt{R} package \texttt{SpatialEpi}.

\appendix

\section*{Appendix A. Proof of Proposition 3.1}

From the construction of the $\bs{\alpha}$- and $\bs{\gamma}$-steps, it is obvious that $\phi(\bs{\alpha}^{(t)}, \bs{\beta}^{(t)}, \bs{\gamma}^{(t)}) 
	\geq \phi(\bs{\alpha}^{(t+1)}, \bs{\beta}^{(t)}, \bs{\gamma}^{(t)})$ and $\phi(\bs{\alpha}^{(t+1)}, \bs{\beta}^{(t+1)}, \bs{\gamma}^{(t)})
	\geq \phi(\bs{\alpha}^{(t+1)}, \bs{\beta}^{(t+1)}, \bs{\gamma}^{(t+1)})$. 
%\sout{For the descent property of the $\bs{\beta}$-step, we verify a more generic lemma. It indicates such a property is satisfied by the modified local quadratic approximation algorithm  for a smooth convex loss function combined with a convex penalty function that is possibly non-differentiable.}
To guarantee $$\phi(\bs{\alpha}^{(t+1)}, \bs{\beta}^{(t)}, \bs{\gamma}^{(t)}) \geq \phi(\bs{\alpha}^{(t+1)}, \bs{\beta}^{(t+1)}, \bs{\gamma}^{(t)}),$$ we verify a more general statement. Lemma 1 indicates that if an objective function $\psi(\cdot)$ consists of a smooth convex loss function $l(\cdot)$ plus a convex (and possibly non-differentiable) penalty $P(\cdot)$, one can descend $\psi$ by minimizing a surrogate function of $\psi$ in which the loss part $l$ is replaced by the local quadratic approximation of $l$. 

\begin{lemma}\label{lemma:1}
Let  $l(\bs{t})$ be a twicely-differentiable convex function and $P(\bs{t})$ be a convex function that is not necessarily differentiable. Define $\psi(\bs{t}) := l(\bs{t}) + P(\bs{t})$. Let $\bs{t}^0$ be a fixed value in the domain of $\psi$. 
Define 
\[
\TD{l}(\bs{t};\bs{t}^0) := l(\bs{t}^0) + \nabla_{\bs{t}} l(\bs{t}^0)^T (\bs{t} - \bs{t}^0) + \frac{1}{2} (\bs{t} - \bs{t}^0)^T \nabla_{\bs{t}\bs{t}}^2 l(\bs{t}^0) (\bs{t} - \bs{t}^0), 
~~~~ \TD{\psi}(\bs{t};\bs{t}^0) := \TD{l}(\bs{t};\bs{t}^0) + P(\bs{t}),
\]
and $\bs{t}^* := \argmin_{\bs{t}} \TD{\psi}(\bs{t};\bs{t}^0)$. 
Consider
\[
\bs{t}^1 := h^*\bs{t}^* + (1-h^*)\bs{t}^0, 
\mbox{~~where~}
h^* := \argmin_{h \in [0,1]}  \psi\left(h\bs{t}^* + (1-h)\bs{t}^0\right).
\] 
If $\bs{t}^0$ is not a minimizer of $\psi(\bs{t})$ (i.e., $\psi(\bs{t}^0) < \psi(\bs{t}^*)$), then such $h^*$ exists and $\psi(\bs{t}^1) < \psi(\bs{t}^0)$.
\end{lemma}

The proof is essentially the same as the proof of Proposition 1 in \cite{Lee2016d}. They assumed $P(\bs{t}) = \lambda\|\bs{t}\|_1$ for some $\lambda$, which can be easily extended to a general convex function $P(\cdot)$.  
\begin{proof}
Note that $\psi(\bs{t})$ and $\TD{\psi}(\bs{t};\bs{t}^0)$ have the same penalty function, and $\nabla_{\bs{t}} l(\bs{t}) = \nabla_{\bs{t}} \TD{l}(\bs{t}^0;\bs{t}^0)$. In addition,  $\psi(\bs{t})$ and $\TD{\psi}(\bs{t};\bs{t}^0)$ have the same Karush-Kuhn-Tucker first-order optimality conditions at $\bs{t} = \bs{t}^0$. Then the assumption  $\bs{t}^0 \neq \argmin_{\bs{t}} \psi(\bs{t})$ is equivalent to $\bs{t}^0 \neq \argmin_{\bs{t}} \TD{\psi}(\bs{t};\bs{t}^0)$. Consequently, $\TD{\psi}(\bs{t}^*;\bs{t}^0) < \TD{\psi}(\bs{t}^0;\bs{t}^0)$. Now let $h \in (0,1]$ and $\bs{t}^h = h\bs{t}^* + (1-h)\bs{t}^0$. The convexity of  $\TD{\psi}(\cdot;\bs{t}^0)$ implies
\[
\TD{\psi}( \bs{t}^h ;\bs{t}^0) \leq 
	h \TD{\psi}( \bs{t}^* ;\bs{t}^0) + 
	(1-h) \TD{\psi}( \bs{t}^0 ;\bs{t}^0),
\]
which yields
\BEQ%\label{eqn:proof1}
\nonumber
\frac{\TD{\psi}( \bs{t}^h ;\bs{t}^0) - \TD{\psi}( \bs{t}^0 ;\bs{t}^0)}
	{h} \leq \TD{\psi}( \bs{t}^* ;\bs{t}^0) - \TD{\psi}( \bs{t}^0 ;\bs{t}^0) < 0.
\EEQ
Furthermore, 
\BEA
&& \frac{\psi(\bs{t}^h) - \psi(\bs{t}^0)}{h} \NN \\
&=& \frac{\TD{\psi}( \bs{t}^h ;\bs{t}^0) - \TD{\psi}( \bs{t}^0 ;\bs{t}^0)}{h}
	- \frac{\TD{l}( \bs{t}^h ;\bs{t}^0) - \TD{l}( \bs{t}^0 ;\bs{t}^0)}{h}
	+ \frac{l(\bs{t}^h) - l(\bs{t}^0)}{h}
	\NN \\
&\leq& \TD{\psi}( \bs{t}^* ;\bs{t}^0) - \TD{\psi}( \bs{t}^0 ;\bs{t}^0)
	- \frac{\TD{l}( \bs{t}^h ;\bs{t}^0) - \TD{l}( \bs{t}^0 ;\bs{t}^0)}{h}
	+ \frac{l(\bs{t}^h) - l(\bs{t}^0)}{h}. \NN
\EEA
As $h \rightarrow 0^+$, $\TD{\psi}( \bs{t}^* ;\bs{t}^0) - \TD{\psi}( \bs{t}^0 ;\bs{t}^0)
	- \{\TD{l}( \bs{t}^h ;\bs{t}^0) - \TD{l}( \bs{t}^0 ;\bs{t}^0)\}/h$
 converges to $-\nabla_{\bs{t}} \TD{l}(\bs{t}^0;\bs{t}^0) + \nabla_{\bs{t}} l(\bs{t})$, which vanish to zero by the construction of $\TD{l}$. Therefore, we have 
\[
\limsup_{h \rightarrow 0^+} \frac{\psi(\bs{t}^h) - \psi(\bs{t}^0)}{h} \leq
	\TD{\psi}( \bs{t}^* ;\bs{t}^0) - \TD{\psi}( \bs{t}^0 ;\bs{t}^0) < 0.
\]
Therefore, there exists at least one $h \in (0, 1]$ such that $\psi(\bs{t}^h) < \psi(\bs{t}^0)$. $\psi(\bs{t}^1) < \psi(\bs{t}^0)$ subsequently according to the construction of $\bs{t}^1$.
\end{proof}

\section*{Appendix B. Modified objective function adjusted for weights}
Denote the final weight as $w_{ij}$ for $(ij)$-th subject. The modified objective function $\phi$ is defined by
\BEQ\label{eqn:objFnW} 
	\phi^w(\bs{\alpha}, \bs{\beta}, \bs{\gamma}) = 
 	-{\rm loglik}^w (\bs{\alpha}, \bs{\beta}, \bs{\gamma}) + 
	P_{\lambda_1}(\bs{\beta}) +  Q_{\lambda_2}^w(\bs{\gamma}),
\EEQ 
where 
\begin{eqnarray*}
{\rm loglik}^w (\bs{\alpha}, \bs{\beta}, \bs{\gamma}) 
&=&
	\frac{1}{w_{\cdot\cdot}} \sum_{i=1}^K \sum_{j=1}^{n_i} w_{ij} I(R_{ij} = 1)\cdot 
	 \bigg[
	 \log\{1 + \exp({\bs Z}_{ij}^T \bs{\alpha}_1   +  {\bs X}_i^T \bs{\alpha}_2
	 \NN \\
&& \qquad + \beta_i + \gamma_i)\} 
	 - Y_{ij} ({\bs Z}_{ij}^T \bs{\alpha}_1 + 
	 {\bs X}_i^T \bs{\alpha}_2 + \beta_i + \gamma_i) \bigg], 
\end{eqnarray*}
and
\[
Q_{\lambda_2}^w(\bs{\gamma}) = \frac{1}{w_{\cdot \cdot}} \sum_{i=1}^K w_{i \cdot} q_{\lambda_2}(\gamma_i),
\]
with $w_{i \cdot} = \sum_j w_{ij} I(R_{ij} = 1)$ and $w_{\cdot \cdot} = \sum_i\sum_j w_{ij} I(R_{ij} = 1)$. The case of complete data can be understood as $w_{ij}=1$ and $R_{ij}=1$ for all $i$ and $j$.
Algorithm 1 in Web Supplementary Materials describes the   {alternating minimization}   algorithm adjusted for the weight. The result of Proposition \ref{prop:1} remains the same as long as the dataset contains at least one obese and non-obese subject observed for all locations. The modified BIC reflecting the weight is
\[
{\rm BIC}^{w*}(\lambda_1, \lambda_2) = - 2 w_{\cdot\cdot} \cdot {\rm loglik}^w(\WH{\bs{\alpha}}, \WH{\bs{\beta}}, \WH{\bs{\gamma}}) + {\rm DF} \cdot \left( \log w_{\cdot\cdot} + 1 \right),
\]
where ${\rm loglik}^w(\WH{\bs{\alpha}}, \WH{\bs{\beta}}, \WH{\bs{\gamma}})$ is given in  \eqref{eqn:objFnW} and ${\rm DF}$ is given in (6) in Section 3.4.

\section*{Appendix C. Optimization algorithm}

Algorithm \ref{algo:1} describes the proposed alternating minimization algorithm solving \eqref{eqn:objFnW}.

\begin{algorithm}[h!]
\caption{An alternating minimization algorithm for \eqref{eqn:objFnW}}\label{algo:1}
\begin{algorithmic}
%\Procedure{MyProcedure}{}
\State {\bf require:} Arrays $\{Y_{ij}\}$, $\{\bs{Q}_{ij}\}$, $\{R_{ij}\}$ and $\{w_{ij}\}$, $i=1,\ldots,n_i$, $j=1,\ldots,K$, array $\{\rho_{i_1, i_2}\}$, scalar $\lambda_1$ and scalar $\lambda_2$, tolerance level $\epsilon = 10^{-6}$

\State {\bf initialize} $\bs{\alpha}^{(0)}$, $\bs{\beta}^{(0)}$, $\bs{\gamma}^{(0)}$, $\phi^{(0)} = \phi(\bs{\alpha}^{(0)}, \bs{\beta}^{(0)}, \bs{\gamma}^{(0)})$
\State define $w_{i \cdot} \gets \sum_j w_{ij} I(R_{ij} = 1)$ and $w_{\cdot \cdot} \gets \sum_i\sum_j w_{ij} I(R_{ij} = 1)$

\While{ $\frac{|\phi^{(t+1)} - \phi^{(t)}|}{\max\{1, |\phi^{(t)}|\}} > \epsilon$}

\State
\State \underline{(1. Updating $\bs{\alpha}$)}
\State 1-1. $\mu_{ij}^{(t)} \gets \beta_i^{(t)} + \gamma_i^{(t)}$ for $j=1, \ldots, n_i$, $i=1,\ldots,K$. 
\State 1-2. Run a logistic regression, without intercept, for $N=\sum_{i=1}^K n_i$ individuals with response $\{Y_{ij}\}$, predictor $\{\bs{Q}_{ij}\}$, offset $\{\mu_{ij}^{(t)}\}$ and weight $\{w_{ij}\}$. 
\State 1-3. Assign the results from Steps 1-1/1-2  to $\bs{\alpha}_i^{(t+1)}$.

\State
\State \underline{(2. Updating $\bs{\beta}$)}
\State 2-1. $\theta_{ij}^{(t)} \gets \bs{Q}_{ij}^T \bs{\alpha}^{(t+1)} + \gamma_i^{(t)}$ for $j=1, \ldots, n_i$, $i=1,\ldots,K$. 
\State 2-2. $a_i^{(t)} \gets \sum_{j=1}^{n_i} w_{ij} \left[  
	\frac{ \exp\left( \beta_i^{(t)} + \theta_{ij}^{(t)} \right)}
	{ \left\{ 1 + \exp\left( \beta_i^{(t)} + \theta_{ij}^{(t)} \right) \right\}^2 }
	\right]$ for $i=1,\ldots,K$.
\State 2-3. $b_i^{(t)} \gets \beta_i^{(t)} - \frac{1}{a_i^{(t)}} \sum_{j=1}^{n_i} w_{ij} \left[  
	\frac{ \exp\left( \beta_i^{(t)} + \theta_{ij}^{(t)} \right)}
	{ 1 + \exp\left( \beta_i^{(t)} + \theta_{ij}^{(t)} \right) }
	- Y_{ij} \right]$ for $i=1,\ldots,K$.
\State 2-4. Solve %the following least square with generalized lasso penalty and identity design matrix to obtain $\bs{\beta}^*$:
\[
\TD{\bs{\beta}} \gets \argmin_{\bs{\beta} \in \MBR^K} \left[ \frac{1}{2 w_{\cdot\cdot}} \sum_{i=1}^K  a_i^{(t)} \left(\beta_i - b_i^{(t)} \right)^2
	+ \lambda_1  \sum_{i_1 < i_2} \rho_{i_1,i_2} |\beta_{i_1} - \beta_{i_2}| \right].
\]
\State 2-5. If $\phi(\bs{\alpha}^{(t+1)}, \TD{\bs{\beta}}, \bs{\gamma}^{(t)}) \leq \phi(\bs{\alpha}^{(t+1)}, \bs{\beta}^{(t)}, \bs{\gamma}^{(t)})$, then $\bs{\beta}^{(t+1)} \gets \TD{\bs{\beta}}$. Otherwise, $\bs{\beta}^{(t+1)} \gets \TD{h} \TD{\bs{\beta}} + (1 - \TD{h})\bs{\beta}^{(t)}$, where
\[
\TD{h} = \argmin_{h \in [0,1]}  \phi\left(\bs{\alpha}^{(t+1)},
	 h\TD{\bs{\beta}} + (1-h)\bs{\beta}^{(t)}, \bs{\gamma}^{(t)}\right).
\]
\State
\State \underline{(3. Updating $\bs{\gamma}$)}
\State 3-1. $\nu_{ij}^{(t)} \gets \bs{Q}_{ij}^T \bs{\alpha}^{(t+1)} + \beta_i^{(t+1)}$ for $j=1, \ldots, n_i$, $i=1,\ldots,K$. 
\State 3-2. For $i=1,\ldots,K$ :
\[
\gamma_i^{(t+1)} \gets \argmin_{\gamma} \left[ \sum_{j=1}^{n_i} w_{ij} \left[
	\log \left\{1 + \exp\left(\gamma + \nu_{ij}^{(t)}\right) \right\} 
	- Y_{ij} \left(\gamma + \nu_{ij}^{(t)}\right) \right]
	+ w_{i \cdot} q_{\lambda_2}(\gamma) \right].
\]
%	\For {$i = 1, \ldots, K$}
%	\State 3-2-1. Define $l_i (\gamma) = \sum_{j=1}^{n_i} w_{ij} \left[
%	\log \left\{1 + \exp\left(\gamma + \nu_{ij}^{(t)}\right) \right\} 
%	- Y_{ij} \left(\gamma + \nu_{ij}^{(t)}\right) \right]$.
%	\State 3-2-2. Define $\phi_i (\gamma) = l_i (\gamma) +  w_{i \cdot} q_{\lambda_2}(\gamma)$.
%	\State 3-2-3. Let 
%	\State initial grid search on $\{t_1, \ldots, t_T \} \subseteq [-\lambda_2, \lambda_2]$ and run a univariate function optimizer with initial point as $\argmin_{\gamma \in \{\TD{\gamma}, t_1, \ldots, t_T \}} \phi_i (\gamma)$.
%	\EndFor

\State 4.  $\phi^{(t+1)} \gets \phi(\bs{\alpha}^{(t+1)}, \bs{\beta}^{(t+1)}, \bs{\gamma}^{(t+1)})$

\State
\EndWhile
\State {\bf return} $(\bs{\alpha}^{(t+1)}, \bs{\beta}^{(t+1)}, \bs{\gamma}^{(t+1)})$
%\EndProcedure
\end{algorithmic}
\end{algorithm}

%\bibliography{library_YG.bib}

\begin{thebibliography}{}

\bibitem[Blondin et~al., 2016]{Blondin2016}
Blondin, K.~J., Giles, C.~M., Cradock, A.~L., Gortmaker, S.~L., and Long, M.~W.
  (2016).
\newblock {US States' childhood obesity surveillance practices and
  recommendations for improving them, 2014–2015}.
\newblock {\em Preventing Chronic Disease}, 13:160060.

\bibitem[CDC, 2016]{CDC2016}
CDC (2016).
\newblock {Behavioral risk factor surveillance system, comparability of data
  BRFSS 2015 (Version {\#}1—Revised: June 2016)}.
\newblock Technical report.

\bibitem[Davila-Payan et~al., 2015]{Davila-Payan2015}
Davila-Payan, C., DeGuzman, M., Johnson, K., Serban, N., and Swann, J. (2015).
\newblock {Estimating prevalence of overweight or obese children and
  adolescents in small geographic areas using publicly available data}.
\newblock {\em Preventing Chronic Disease}, 12:E32.

\bibitem[ESRI, 2012]{ESRI2012}
ESRI (2012).
\newblock {\em {Environmental Systems Research Institute (ESRI): tapestry
  segmentation reference guide}}.
\newblock CA: Redlands.

\bibitem[Farrington et~al., 1996]{Farrington1996}
Farrington, C.~P., Andrews, N.~J., Beale, A.~D., and Catchpole, M.~A. (1996).
\newblock {A statistical algorithm for the early detection of outbreaks of
  infectious disease}.
\newblock {\em Journal of the Royal Statistical Society. Series A (Statistics
  in Society)}, 159(3):547.

\bibitem[Flood et~al., 2015]{Flood2015}
Flood, T.~L., Zhao, Y.-Q., Tomayko, E.~J., Tandias, A., Carrel, A.~L., and
  Hanrahan, L.~P. (2015).
\newblock {Electronic health records and community health surveillance of
  childhood obesity}.
\newblock {\em American Journal of Preventive Medicine}, 48(2):234--240.

\bibitem[Friedman et~al., 2013]{Friedman2013}
Friedman, D.~J., Parrish, R.~G., and Ross, D.~A. (2013).
\newblock {Electronic health records and US public health: Current realities
  and future promise}.
\newblock {\em American Journal of Public Health}, 103(9):1560--1567.

\bibitem[Friedman et~al., 2010]{Friedman2010a}
Friedman, J., Hastie, T., and Tibshirani, R. (2010).
\newblock {Regularization paths for generalized linear models via coordinate
  descent}.
\newblock {\em Journal of Statistical Software}, 33(1):1--22.

\bibitem[Ghosh et~al., 1999]{Ghosh1999}
Ghosh, M., Natarajan, K., Waller, L.~A., and Kim, D. (1999).
\newblock {Hierarchical Bayes {\{}GLMs{\}} for the analysis of spatial data: An
  application to disease mapping}.
\newblock {\em Journal of Statistical Planning and Inference}, 75(2):305--318.

\bibitem[Greenland, 2004]{Greenland2004}
Greenland, S. (2004).
\newblock {The impact of prior distributions for uncontrolled confounding and
  response bias}.
\newblock {\em Journal of the American Statistical Association},
  98(461):47--54.

\bibitem[Guilbert et~al., 2012]{guilbert2012}
Guilbert, T.~W., Arndt, B., Temte, J., Adams, A., Buckingham, W., Tandias, A.,
  Tomasallo, C., Anderson, H.~A., and Hanrahan, L.~P. (2012).
\newblock {The theory and application of UW ehealth-PHINEX, a clinical
  electronic health record-public health information exchange}.
\newblock {\em WMJ}, 111(3):124--33.

\bibitem[Hoelscher et~al., 2017]{Hoelscher2017}
Hoelscher, D.~M., Ranjit, N., and P{\'{e}}rez, A. (2017).
\newblock {Surveillance systems to track and evaluate obesity prevention
  efforts}.
\newblock {\em Annual Review of Public Health}, 38(1):187--214.

\bibitem[Hyun et~al., 2016]{Hyun2016}
Hyun, S., G'sell, M., and Tibshirani, R.~J. (2016).
\newblock {Exact post-selection inference for changepoint detection and other
  generalized lasso problems}.
\newblock {\em arXiv preprint arXiv: 1606.03552}.

\bibitem[Jung, 2009]{Jung2009}
Jung, I. (2009).
\newblock {A generalized linear models approach to spatial scan statistics for
  covariate adjustment}.
\newblock {\em Statistics in Medicine}, 28(7):1131--1143.

\bibitem[Kafadar and Stroup, 1992]{Kafadar1992}
Kafadar, K. and Stroup, D.~F. (1992).
\newblock {Analysis of aberrations in public health surveillance data:
  Estimating variances on correlated samples}.
\newblock {\em Statistics in Medicine}, 11(12):1551--1568.

\bibitem[Kim et~al., 2009]{Kim2009}
Kim, S.-J., Koh, K., Boyd, S., and Gorinevsky, D. (2009).
\newblock {l1 Trend Filtering}.
\newblock {\em SIAM Review}, 51(2):339--360.

\bibitem[Kulldorff, 1997]{Kulldorff1997}
Kulldorff, M. (1997).
\newblock {A spatial scan statistic}.
\newblock {\em Communications in Statistics - Theory and Methods},
  26(6):1481--1496.

\bibitem[Kulldorff and Nagarwalla, 1995]{Kulldorff1995}
Kulldorff, M. and Nagarwalla, N. (1995).
\newblock {Spatial disease clusters: Detection and inference}.
\newblock {\em Statistics in Medicine}, 14(8):799--810.

\bibitem[Lee, 2013]{Lee2013a}
Lee, D. (2013).
\newblock {CARBayes: An R Package for Bayesian spatial modeling with
  conditional autoregressive priors}.
\newblock {\em Journal of Statistical Software}, 55(13):1--24.

\bibitem[Lee et~al., 2016]{Lee2016d}
Lee, S., Kwon, S., and Kim, Y. (2016).
\newblock {A modified local quadratic approximation algorithm for penalized
  optimization problems}.
\newblock {\em Computational Statistics and Data Analysis}, 94:275--286.

\bibitem[Little and Rubin, 2014]{little2014statistical}
Little, R.~J. and Rubin, D.~B. (2014).
\newblock {\em {Statistical Analysis with Missing Data}}.
\newblock John Wiley {\&} Sons.

\bibitem[Longjohn et~al., 2010]{Longjohn2010}
Longjohn, M., Sheon, A.~R., Card-Higginson, P., Nader, P.~R., and Mason, M.
  (2010).
\newblock {Learning from State surveillance of childhood obesity}.
\newblock {\em Health Affairs}, 29(3):463--472.

\bibitem[Maiti et~al., 2016]{Maiti2016}
Maiti, T., Sinha, S., and Zhong, P.-S. (2016).
\newblock {Functional mixed effects model for small area estimation}.
\newblock {\em Scandinavian Journal of Statistics, Theory and Applications},
  43(3):886--903.

\bibitem[{Mass. Dep. Public Health}, 2014]{Mass2014}
{Mass. Dep. Public Health} (2014).
\newblock {BMI screening guidelines for schools}.
\newblock Technical report, Mass. Dep. Public Health, Boston, MA.

\bibitem[Mercer et~al., 2015]{Mercer2015}
Mercer, L.~D., Wakefield, J., Pantazis, A., Lutambi, A.~M., Masanja, H., and
  Clark, S. (2015).
\newblock {Space–time smoothing of complex survey data: Small area estimation
  for child mortality}.
\newblock {\em The Annals of Applied Statistics}, 9(4):1889--1905.

\bibitem[Nathan and Adams, 1989]{Nathan1989}
Nathan, R.~P. and Adams, C.~F. (1989).
\newblock {Four Perspectives on Urban Hardship}.
\newblock {\em Political Science Quarterly}, 104(3):483.

\bibitem[Panczak et~al., 2016]{Panczak2016}
Panczak, R., Held, L., Moser, A., Jones, P.~A., R{\"{u}}hli, F.~J., and Staub,
  K. (2016).
\newblock {Finding big shots: Small-area mapping and spatial modeling of
  obesity among Swiss male conscripts}.
\newblock {\em BMC Obesity}, 3(1):1--12.

\bibitem[Pascutto et~al., 2000]{Pascutto2000}
Pascutto, C., Wakefield, J.~C., Best, N.~G., Richardson, S., Bernardinelli, L.,
  Staines, A., and Elliott, P. (2000).
\newblock {Statistical issues in the analysis of disease mapping data}.
\newblock {\em Statistics in Medicine}, 19(17-18):2493--519.

\bibitem[She and Owen, 2011]{She2011}
She, Y. and Owen, A.~B. (2011).
\newblock {Outlier detection using nonconvex penalized regression}.
\newblock {\em Journal of the American Statistical Association},
  106(494):626--639.

\bibitem[Snyder et~al., 2018]{Snyder2018}
Snyder, J.~W., Bauer, C.~R., Beaulieu-Jones, B.~K., Pendergrass, S.~A., Lavage,
  D.~R., and Moore, J.~H. (2018).
\newblock {Characterizing and managing missing structured data in electronic
  health records: Data analysis}.
\newblock {\em JMIR Medical Informatics}, 6(1):e11.

\bibitem[Taylor and Tibshirani, 2018]{Taylor2016}
Taylor, J. and Tibshirani, R. (2018).
\newblock {Post-selection inference for l1-penalized likelihood models}.
\newblock {\em Canadian Journal of Statistics}, 46(1):41--61.

\bibitem[Tibshirani et~al., 2005]{Tibshirani2005}
Tibshirani, R., Saunders, M., Rosset, S., Zhu, J., and Knight, K. (2005).
\newblock {Sparsity and smoothness via the fused lasso}.
\newblock {\em Journal of the Royal Statistical Society. Series B: Statistical
  Methodology}, 67:91--108.

\bibitem[Tibshirani and Taylor, 2011]{Tibshirani2011a}
Tibshirani, R.~J. and Taylor, J. (2011).
\newblock {The solution path of the generalized lasso}.
\newblock {\em The Annals of Statistics}, 39(3):1335--1371.

\bibitem[Ugarte et~al., 2010]{Ugarte2010}
Ugarte, M.~D., Goicoa, T., and Militino, A.~F. (2010).
\newblock {Spatio-temporal modeling of mortality risks using penalized
  splines}.
\newblock {\em Environmetrics}, 21(3-4):270--289.

\bibitem[Waller et~al., 1997]{Waller1997}
Waller, L.~A., Carlin, B.~P., Xia, H., and Gelfand, A.~E. (1997).
\newblock {Hierarchical spatio-temporal mapping of disease rates}.
\newblock {\em Journal of the American Statistical Association},
  92(438):607--617.

\bibitem[Yu et~al., 2015]{Yu2015}
Yu, D., Won, J.-H., Lee, T., Lim, J., and Yoon, S. (2015).
\newblock {High-dimensional fused lasso regression using
  majorization-minimization and parallel processing}.
\newblock {\em Journal of Computational and Graphical Statistics},
  24(1):121--153.

\bibitem[Zhang et~al., 2011]{Zhang2011}
Zhang, Z., Zhang, L., Penman, A., and May, W. (2011).
\newblock {Using small-area estimation method to calculate county-level
  prevalence of obesity in Mississippi, 2007-2009}.
\newblock {\em Preventing Chronic Disease}, 8(4):A85.

\bibitem[Zhao et~al., 2011]{Zhao2011}
Zhao, Y., Zeng, D., Herring, A.~H., Ising, A., Waller, A., Richardson, D., and
  Kosorok, M.~R. (2011).
\newblock {Detecting disease outbreaks using local spatiotemporal methods}.
\newblock {\em Biometrics}, 67(4):1508--1517.

\bibitem[Zou et~al., 2007]{Zou2007}
Zou, H., Hastie, T., and Tibshirani, R. (2007).
\newblock {On the ``degrees of freedom'' of the lasso}.
\newblock {\em The Annals of Statistics}, 35(5):2173--2192.

\end{thebibliography}

\end{document}